\documentclass[iop,numberedappendix,appendixfloats]{emulateapj}
\usepackage{epstopdf}
\usepackage{graphics}
\usepackage{natbib}

\newcommand{\Msol}{\mbox{M$_{\odot}$}}
\newcommand{\msun}{\mbox{$\rm M_{\odot}$}}

\newcommand{\lsun}{\mbox{L$_{\odot}$}}
\newcommand{\kms}{\mbox{km s$^{-1}$}}

\newcommand{\etal}[1]{{ et al.}~}
\def\kms{\ifmmode \hbox{km~s}^{-1}\else km~s$^{-1}$\fi}
\def\etal {{\it et al.}}
\def\deg      {{\ifmmode^\circ\else$^\circ$\fi} } %%% Overwrites TeX \deg
\def\degg      {{\ifmmode^\circ\else$^\circ$\fi}} %%% Overwrites TeX \deg

\def\h2     {H$_2$}

\def\arcsec{\hbox{$^{\prime\prime~}$}}

 \let\h=\eta   
\let\l=\lambda \let\m=\mu    
 \let\t=\tau

\def\ba{\begin{array}} \def\ea{\end{array}}
\def \be {\begin{eqnarray}} \def\ee{\end{eqnarray}}
\def \bei {\begin{itemize}} \def \eei {\end{itemize}}

\shorttitle{Arp 220 nuclei}
\shortauthors{Scoville \etal}

\begin{document}

\title{ALMA Resolves the Nuclear Disks of Arp 220}

\author{ 
Nick Scoville\altaffilmark{1},
Lena Murchikova\altaffilmark{1},  
Fabian Walter\altaffilmark{2}, 
Catherine Vlahakis\altaffilmark{3},
Jin Koda\altaffilmark{15}, 
Paul Vanden Bout\altaffilmark{10},
Joshua Barnes\altaffilmark{5,4}, 
Lars Hernquist\altaffilmark{6},   
Kartik Sheth\altaffilmark{20}, 
Min Yun\altaffilmark{4},
David Sanders\altaffilmark{5}, 
Lee Armus,\altaffilmark{16},
Pierre Cox,\altaffilmark{17,18},
Todd Thompson\altaffilmark{13,14}, 
Brant Robertson\altaffilmark{11}, 
Laura Zschaechner\altaffilmark{2},
Linda Tacconi\altaffilmark{9}, 
Paul Torrey,\altaffilmark{19,7},
Christopher C. Hayward\altaffilmark{7}, 
Reinhard Genzel\altaffilmark{9}, 
Phil Hopkins\altaffilmark{1},  
Paul van der Werf\altaffilmark{12},  
Roberto Decarli\altaffilmark{2}}
 
%\altaffiltext{}{}
\altaffiltext{1}{California Institute of Technology, MC 249-17, 1200 East California Boulevard, Pasadena, CA 91125}
\altaffiltext{2}{Max-Planck-Institut fur Astronomie, Konigstuhl 17, D-69117 Heidelberg, Germany}
\altaffiltext{3}{National Radio Astronomy Observatory, 520 Edgemont Road, Charlottesville, VA 22901, USA}
\altaffiltext{4}{Yukawa Institute for Theoretical Physics, Kyoto University, Sakyo-ku, Kyoto 606-8502, Japan}
\altaffiltext{5}{Institute for Astronomy, 2680 Woodlawn Dr., University of Hawaii, Honolulu, Hawaii, 96822}
\altaffiltext{6}{Harvard-Smithsonian Center for Astrophysics, 60 Garden Street, Cambridge, MA 02138, USA}
\altaffiltext{7}{TAPIR 350-17, California Institute of Technology, 1200 E. California Boulevard, Pasadena, CA 91125}
\altaffiltext{8}{Department of Astronomy, University of Massachusetts, Amherst, MA 01003}
\altaffiltext{9}{Max-Planck-Institut fur extraterrestrische Physik (MPE), Giessenbachstr., D-85748 Garching, Germany}
\altaffiltext{10}{National Radio Astronomy Observatory, 520 Edgemont Road, Charlottesville, VA 22901, USA}
\altaffiltext{11}{Department of Astronomy and Astrophysics, University of California, Santa Cruz, 1156 High Street, Santa Cruz, CA 95064 }
\altaffiltext{12}{Leiden Observatory, Leiden University, P.O. Box 9513, NL-2300 RA Leiden, The Netherlands}
\altaffiltext{13}{Department of Astronomy, The Ohio State University, 140 West 18th Avenue, Columbus, OH 43210, USA}
\altaffiltext{14}{Center for Cosmology and AstroParticle Physics, The Ohio State University, 191 West Woodruff Avenue, Columbus, OH 43210, USA}
\altaffiltext{15}{Department of Physics \& Astronomy, Stony Brook University, Stony Brook, NY 11794, USA}
\altaffiltext{16}{Infrared Processing and Analysis Center, California Institute of Technology, 1200 E. California Boulevard, Pasadena, CA 91125, USA}
\altaffiltext{17}{Joint ALMA Observatory, Alonso de C$\acute{o}$rdova 3107, Vitacura, Santiago, Chile}
\altaffiltext{}{}\altaffiltext{}{}\altaffiltext{}{}
\altaffiltext{18}{European Southern Observatory, Alonso de C$\acute{o}$rdova 3107, Vitacura, Santiago, Chile}
\altaffiltext{19}{Department of Physics, Kavli Institute for Astrophysics and Space Research, Massachusetts Institute of Technology, Cambridge, MA 02139, USA}
\altaffiltext{20}{NASA Headquarters, 300 E Street SW, Washington DC 20546}
\altaffiltext{}{}\altaffiltext{}{}\altaffiltext{}{}\altaffiltext{}{}\altaffiltext{}{}\altaffiltext{}{}\altaffiltext{}{}\altaffiltext{}{}\altaffiltext{}{}\altaffiltext{}{}\altaffiltext{}{}\altaffiltext{}{}
\altaffiltext{}{}\altaffiltext{}{}\altaffiltext{}{}
\altaffiltext{}{}\altaffiltext{}{}\altaffiltext{}{}
\altaffiltext{}{}\altaffiltext{}{}\altaffiltext{}{}\altaffiltext{}{}\altaffiltext{}{}\altaffiltext{}{}\altaffiltext{}{}\altaffiltext{}{}\altaffiltext{}{}
\altaffiltext{}{}\altaffiltext{}{}\altaffiltext{}{}\altaffiltext{}{}
\altaffiltext{}{}
%\maketitle
%~~~~~~~~~~~~~~~~~~~~~~~~~~~~~~~~~~~~~~~~~~~~Accepted ApJ 2/25/13
\altaffiltext{}{}

\vfill
\eject

\begin{abstract}
We present 90 mas (37 pc) resolution ALMA imaging of Arp 220 in the CO (1-0) line and  
continuum at $\lambda = 2.6$ mm. The internal gas distribution and kinematics of 
both galactic nuclei are well-resolved for the first time. In the West 
nucleus, the major gas and dust emission extends out to 0.2\arcsec radius (74 pc); 
the central resolution element  shows a strong peak in the dust emission but a 
factor 3 dip in the CO line emission. In this nucleus, the dust is apparently 
optically thick ($\tau_{\rm 2.6mm} \sim1$)  at $\lambda = 2.6$ mm with a 
dust brightness temperature $\sim147$ K. The column of ISM at this 
nucleus is $\rm N_{H2} \geq 2\times10^{26}$ cm$^{-2}$, corresponding to  
$\sim$900 gr cm$^{-2}$. The East nucleus is more 
elongated with radial extent 0.3\arcsec or $\sim111$ pc.   The derived kinematics 
of the nuclear disks provide a good fit to the line profiles, 
yielding the emissivity distributions, the rotation curves and velocity dispersions.  
In the West nucleus, there is evidence of a central Keplerian component requiring a 
central mass of $8\times10^8$ \msun.  The intrinsic widths of the emission lines 
are $\Delta \rm v (FWHM)$ = 250 (West) and 120 (East) \kms. Given the very short 
dissipation timescales for turbulence ($\lesssim10^5$ yrs), we suggest that the 
line widths may be due to semi-coherent 
motions within the nuclear disks. The  
symmetry of the nuclear disk structures is impressive -- implying the merger 
timescale is significantly longer than the rotation period of the disks.
\end{abstract}
%The dynamical masses (not including the velocity dispersion component) are $\sim 1.5\times10^9$ \msun within radii 80 (West) and 130 pc (East). 
%In view of the very large central concentration of dust and gas in the central 0.1\arcsec of the West nucleus, the point mass is not necessarily a central black hole. 
\medskip 


\section{Introduction}

%In the era of precision cosmology and vast cosmological simulations consistent implementation of galactic mergers became a crucial element in getting the predictions right. 

Galactic merging is a key process in the early growth and evolution of massive galaxies and in determining their structural morphology. In the era of 
precision cosmology, these processes remain a major uncertainty in understanding the present makeup of the visible universe. 
%At present, merger evolution is only qualitatively well developed. 
Many aspects of the evolution of merging, gas-rich {\it nuclear disks} are poorly constrained and only approximately understood in terms of the physical processes and the hierarchy of their importance.
The radial mass and star formation distributions, the physical conditions (density, temperature and cloud structures) and the 
evolution associated with feedback from starbursts (SBs) and active galactic nuclei (AGN) remain poorly understood. And yet, all of these are vital to predicting the ultimate fate or product of the mergers (e.g. the resulting galactic morphologies) and the mode(s) of star formation and AGN fueling. %Of immediate cosmological importance are: the SB properties -- star formation efficiency, distribution and feedback and the AGN fueling and feedback of processed matter to the circum-galactic environment. 

Ultra-luminous infrared galaxies (ULIRGs) are the most extreme SB galaxies. The first complete sampling of the local universe yielded 22 ULIRGs at z $<$ 0.1 from the IRAS all sky with L$_{1-1000\mu m} > 10^{12}$ \lsun \citep{san88}. At high redshift when the rate of collisions of the galaxies was higher \citep{rod15} and galaxies were more gas-rich, they were much more abundant \citep{lef09,capu07,mag13}. 

Resolved studies of the merging processes  must rely on the sample of local galaxies. The follow-up ground-based optical imaging 
reveals virtually all of the local ULIRGs to be merging galaxies or post merging systems \citep{arm87,san88,san96}. During the merging of gas-rich galaxies, the original ISM (presumably distributed in extended galactic-scale disks) sinks rapidly to the center of the merging system due to dissipation of kinetic energy in the shocked gas and torques generated by the offset stellar and gaseous bars \citep[e.g.][]{bar92,bar96}. The star formation rates in the ULIRGs are typically 10-100 times higher per unit mass of ISM compared to quiescent disk galaxies. The ULIRG-starburst activity is likely driven by  
concentration of gas in nuclear regions and dynamical compression of this gas in supersonic shocks. 

Among the ULIRGs, Arp220 
 is probably the most frequently cited example, having luminosity L$_{IR} = 1.91\times10^{12}$\lsun\citep{san03,arm09}. (Here we adopt a luminosity distance $\rm D_L = 87.9$ Mpc and angular size distance $\rm D_A = 85.0$ Mpc \citep{arm09}.)  Near infrared  imaging shows two galactic nuclei, separated by $1.0$\arcsec or 412 pc \citep{sco98}. CO imaging at 0.5\arcsec resolution uncovers two counter rotating disks with radii $\sim100$ pc and dynamical masses $\sim 2\times 10^9 \Msol$ for each disk \citep{sco97,sak99,sco98,dow07}.
Much of the nuclear ISM is very dense ($>10^{4-5}$ cm$^{-3}$) and at high temperature ($>75$ K) \citep{sak99,dow07,mat09,gre09,ran11,wil14,sco15}.

\begin{deluxetable*}{lcccccccr}
\tabletypesize{\scriptsize}
\tablecaption{\bf{Fluxes} \label{fits}  }
\tablewidth{0pt}
\tablehead{\colhead{Source}  & \colhead{Frequency} & \colhead{$\alpha_{2000}$\tablenotemark{a}} & \colhead{$\delta_{2000}\tablenotemark{a}$} & \colhead{Total Flux}  & \colhead{Peak Flux}    & \colhead{Peak $T_B$}}    
\startdata
\\ 
 \bf{Continuum} & \colhead{(GHz)}  & \colhead{}  &  \colhead{}  & \colhead{(mJy)} &  \colhead{(mJy/beam)} & \colhead{(K)} 
\\   & & & & & & &  \\
        % some of the intensities come from MASK_MEASURE_PLOT
 Arp 220 West &   112.26 &  15:34:57.224 &   23:30:11.48 & 32.1 & 13.1 $\pm$ 0.05 &   147.2 \\ 
Arp 220 East &    112.26 &  15:34:57.286 &   23:30:11.32 & 14.0  & 3.1 $\pm$ 0.04 &    30.5 \\ 
    \\  & & & & & & & \\
 {\bf{Lines}} & & & &   (Jy km s$^{-1}$)   &  (Jy km s$^{-1}$ beam$^{-1}$)  & (K) \\
& & & & & & &\\    
Arp 220 West & CO (1 - 0)  &  15:34:57.224 &   23:30:11.49 & 47.3  & 3.33 $\pm$ 0.01 & 187\tablenotemark{b}  \\ 
Arp 220 East & CO (1 - 0)  &  15:34:57.290 &   23:30:11.34 & 27.8  & 1.78 $\pm$ 0.01 & 175\tablenotemark{b} \\ 
 \\
  \enddata
\tablecomments{Peak position and fluxes obtained from 2d Gaussian fits using the IDL routine CURVEFIT. Gaussian component sizes are listed in Table \ref{sizes}. Uncertainties in the continuum and line fluxes obtained from the uncertainties in the Gaussian component fitting do not include 
calibration uncertainties.}
\tablenotetext{a}{Peak derived from two-dimensional Gaussian fit (Table \ref{sizes}).}
\tablenotetext{b}{Peak brightness temperature from the spectral cube within $\pm 600$ \kms ~of the systemic velocity for the line or from the continuum images of the dust.}

\end{deluxetable*}

Both
nuclei are optically thick at $\lambda < 600 \, \mu$m.  \cite{wil14} derive dust optical depths $\t_{434\mu m}=5.3$ and 1.7 for the West and East nuclei, respectively, implying that for the West nucleus $\t_{1 mm} \sim 1$. These high optical depths imply that the nuclear disk structures are best probed at $\l > 600 \, \m$m. Shorter wavelength observations may not penetrate the outer dust photospheres of the nuclei unless the structures are tilted to the line of sight. The fact than any near infrared radiation can be seen from the nuclei
is a clear indication that the dust is in a disk-like distribution tilted to the line of sight.

%Our previous ALMA imaging of Arp 220 in HCN (4-3) at 0.55\arcsec resolution confirmed the presence of two counter rotating nuclear disks, but did not resolve each internally \citep{sco15}. %Band 9 0.25'' resolution continuum imaging obtained in the same project revealed stronger constraints on the densities of the ISM around the nuclei. 
%The total mass of the ISM was estimated from the dust emission at $2-4\times10^9$ \msun in each nucleus, concentrated within radii $\lesssim70$ pc. This concentration is equivalent to having the entire Milky Way gas mass packed into a single molecular cloud with average density of $10^{5-6} \rm cm^{-3}$ --  a spectacular {\it perturbation} to the normal conditions for star formation.

Here, we present 90 mas (37 pc) resolution ALMA imaging of the inner region of Arp220 in the CO (1-0) line and the 2.6 mm continuum
(108 - 114 GHz, ALMA Band 3), providing excellent resolution and sensitivity for imaging the molecular gas and the long wavelength dust continuum. The East and West nuclei are internally resolved for the first time.  The dust continuum provides an independent and  linear probe of the overall ISM mass \citep{sco15,sco16}.

\section{ALMA Observations and Data Reduction}\label{obs}

%40 antennas
%30 minutes
%1.5 hours including calibrations
%12m antennas only
%max distance between antennas 11.5 km
%ra 15:34:57.217
%dec +23:30:11.44
%average to 40kms
%

These ALMA Cycle 3 long baseline observations in CO (1-0) line 
were obtained in 2015 November for project \#2015.1.00113.S. (We are also scheduled to obtain CO (2-1) and (3-2) high and low resolution imaging but 
those data will probably not be available until the end of 2017.) In view of the major increase in resolution provided by the high resolution CO (1-0) data, publication of this data is important and we proceed here with those preliminary results. 

The observations discussed here were in receiver Band 3; the correlator was configured in the time division
mode (TDM) with 4 spectral windows. Each window had a full bandwidth of 1875 MHz with
1.95 MHz resolution spectral channels.  One window was
configured to observe the redshifted CO (1-0) line at rest frequency of 115.2712 GHz; the
remaining three spectral windows were centered at 113.253 GHz, 103.073 GHz and 101.139 GHz to image the dust continuum emission and $^{13}$CO and 
C$^{18}$O. The latter will be presented in our later publication with the other bands.  

The observations were done in a very extended
configuration with baselines up to 11 km, providing the maximum resolution presently available with ALMA. For this telescope configuration, good flux recovery 
is expected out to scales of
$\sim 0.4$\arcsec but extended emission with spatial size greater than this will be at least partially resolved out. These observations thus probe only the inner nuclei of Arp 220. The data were taken with 40 12 m antennas and the total integration time
was 30 minutes (excluding calibrations).  

Following delivery of data products, the data were re-reduced and imaged
using the Common Astronomy Software Applications package (CASA).
Self-calibration was done to improve the dynamic range.  The images were made with the parameter Robust = 0, 1 and 2; only the Robust = 1 images 
are used here. We cleaned the images with no continuum subtraction, noted the line and absorption free channels and then created a continuum image using those channels. The continuum subtraction was then done in the 
image plane.

Channel averaging over 8 of the original channels was done to reduce noise, resulting in data with velocity resolution of 40 \kms
~without serious compromise relative to the intrinsic line width ($\Delta v_{FWHM} = 100 - 200$ \kms, see Section \ref{modeling}).
The 1$\sigma$ (rms) sensitivities are as follows: an rms noise of $0.6\rm ~mJy ~beam^{-1}$ in 40~\kms ~channels in the
lower sideband at $\sim 110$ GHz) and $ 0.8\rm ~mJy ~beam^{-1}$  in 40 \kms channels in the upper sideband at 114 GHz. 

The velocities given here are $v_{radio} = cz/(1+z) =c (\nu_{rest}-\nu)/ \nu_{rest}$ relative to the LSR (not $v_{opt} = cz = c (\nu_{rest}-\nu)/\nu$). 
The Arp 220 observations were centered on z = 0.018486, implying $cz = v_{opt} = 5542$ \kms and $v_{radio} = 5441$ \kms \citep{san91}. The derived systemic 
velocities of the nuclei are $v_{radio} = 5337$ (West) and 5431 \kms (East) (Table \ref{models}).

Table \ref{fits} lists the measured source fluxes and peak brightness temperatures for both the continuum and the CO (1-0) line 
and Table 2 contains the results of two-dimensional Gaussian fits to each of the sources. The total recovered CO line flux from the two nuclei is $45.4+27.4 = 72.8$ Jy \kms. This is 19\% of the total single dish CO (1-0) line flux measured by \cite{san91} and \cite{sol97}. 

\begin{deluxetable*}{lrllrrccccccr}
\tabletypesize{\scriptsize}

\tablecaption{\bf{Gaussian Source Fits} \label{sizes}  }
\tablewidth{0pt}
\tablehead{  &\multicolumn{4}{c}{\bf{Gaussian Fit }} &\multicolumn{4}{c}{\bf{Deconvolved}}\\
  & \multicolumn{4}{c}{---------------------------------------------------------------} & \multicolumn{4}{c}{-----------------------------------------}\\
\colhead{Source}    &  \colhead{Peak Flux} &  \colhead{Major} & \colhead{Minor} &  \colhead{PA}  &  \colhead{Major} & \colhead{Minor}&  \colhead{PA} &  \colhead{T$_B$}   \\
\colhead{}   &   & \colhead{(\arcsec)} &  \colhead{(\arcsec)} &\colhead{(\deg)}  & \colhead{(\arcsec)} &  \colhead{(\arcsec)} &\colhead{(\deg)} & (K)}\\
\startdata
\\ 
{\bf{112 GHz Continuum}}  & (mJy/beam)\\
\\
Arp 220 West &   11.87 $\pm$ 0.02  & 0.13 $\pm$ 0.01 &   0.12 $\pm$ 0.01 &   50.7 &  0.12 &  0.11 &  64.1 &     167.0 \\ 
Arp 220 East &    2.46 $\pm$ 0.02  & 0.24 $\pm$ 0.02 &   0.15 $\pm$ 0.01 &   49.2 &  0.24 &  0.14 &  50.7 &      33.2 \\ 
   \\
{\bf{CO(1-0) Lines}} &  (Jy km s$^{-1}$)    & \\
&  & & & & & & &\\
Arp 220 West &   2.96 $\pm$ 0.01  & 0.37 $\pm$ 0.01 &   0.33 $\pm$ 0.02 &  164.5 &  0.36 &  0.32 & 162.7 &   - \\ 
Arp 220 East &  1.44 $\pm$ 0.01 &  0.54 $\pm$ 0.02 &   0.25 $\pm$ 0.01 &   46.2 &  0.54 &  0.24 &  46.5 &   - \\ 
\enddata
\tablecomments{Sizes (FWHM) and major axis PA estimates obtained from 2d Gaussian fits using the IDL routine CURVEFIT. The deconvolved sizes were obtained using the IDL routine MAX\_LIKELIHOOD. The uncertainties in the parameters were the formal errors from the Gaussian fitting. For all of the observations the synthesized beam was $0.08 \times 0.10$\arcsec ~at PA = 25.7\deg.}
\end{deluxetable*}

\begin{deluxetable*}{lccrrl}
\tabletypesize{\scriptsize}
%\normaltsize
%\rotate
\tablecaption{\bf{Continuum Fluxes} \label{cont}  }
\tablewidth{0pt}
\tablehead{  \colhead{$\nu_{obs}$} & \colhead{Beam} & \colhead{Total Flux}  & \colhead{Peak Flux}  & \colhead{Peak $T_B$\tablenotemark{a}}  & \colhead{Reference}  \\
  \colhead{(GHz)} &  \colhead{($\arcsec\times  \arcsec$)}  & \colhead{(mJy)} &  \colhead{(mJy/beam)} & \colhead{(K)} }
\startdata
\\
  \multicolumn{2}{c}{\bf{West Nucleus}}\\
 \\
 4.70 & $0.60 \times 0.43$  & 114.6 & 89.5 $\pm$ 0.7  & 1.92$\times10^4$ & \cite{bar15} \\ 
   5.95 & $0.38 \times 0.28$  & 94.3 & 73.3 $\pm$ 1.0  & 2.38$\times10^4$ & \cite{bar15} \\ 
  7.20 & $0.38 \times 0.28$  & 89.5 & 60.4 $\pm$ 0.7  &  1.34$\times10^4$ & \cite{bar15} \\ 
    32.5 & $0.08 \times 0.06$  & 33.4 & 6.5 $\pm$ 0.8  & 1570  & \cite{bar15} \\ 
  112.3  & $0.10 \times 0.08$  & 29.0 &13.1 $\pm$ 0.5  & 121\tablenotemark{b} & Scoville et al 2017 \\ 
 229.4 & $0.30\times0.30$ & 106 & 79.0  $\pm$ 2.0 & 90 &  \cite{dow07} \\ 
341.8 & $0.55 \times 0.40$  & 328 & 261 $\pm$ 2  & 12.4 & \cite{sco15} \\
347.9 & $0.46 \times 0.39$ & 384 & 283 $\pm$ 0.4 & 80 & \cite{mar16} \\
349.6 & $0.25\times 0.21$ & 360 & 360 $\pm$ 50 & 160 & \cite{sak08} \\
  691 & $0. 36 \times 0. 20$  &1810 & 1150 $\pm$ 11 & 40.9 & \cite{wil14} \\
% 698 & $0.32 \times 0.28$  & 3800 & 2750 $\pm$ 20 & 77.0 & \cite{sco15} \\
\\
  \multicolumn{2}{c}{\bf{East Nucleus}}\\
\\
  4.70  & $0.60 \times 0.43$  & 92.4 & 61.8 $\pm$ 0.5 & 1.33$\times10^4$ &  \cite{bar15} \\ 
  5.95  & $0.38 \times 0.28$  & 81.4 & 49.0 $\pm$ 0.7 & 1.59$\times10^4$ &  \cite{bar15} \\ 
  7.20  & $0.38 \times 0.28$  & 73.2 & 36.0 $\pm$ 0.4 & 7980 &  \cite{bar15} \\ 
 32.5  & $0.08 \times 0.06$  & 30.1 & 4.1 $\pm$ 0.5 & 988 &  \cite{bar15} \\ 
  112.3  & $0.10 \times 0.08$  & 16.1 & 3.1 $\pm$ 0.5 & 15\tablenotemark{b} & Scoville et al 2017 \\ 
 341.8 & $0.55 \times 0.40$ & 157 & 111 $\pm$ 3 & 5.3 & \cite{sco15} \\
 347.9 & $0.46 \times 0.39$ & 192 & 115 $\pm$ 0.4 & 15 &  \cite{mar16} \\
 349.6  & $0.25\times 0.21$ &    190 & 190 $\pm$ 50 & 52  & \cite{sak08} \\
  691 & $0. 36 \times 0. 20$  &1510 & 800 $\pm$ 11 & 28.4  & \cite{wil14} \\
% 698 & $0.32 \times 0.28$  & 2150 & 1360 $\pm$ 60 & 38.1  & \cite{sco15} \\
  \enddata
\tablecomments{Includes only continuum data at $<0.6$\arcsec~ resolution. Scoville \etal 2017 is this paper.}
\tablenotetext{a}{Peak brightness temperature in the continuum calculated from $T_B = ( S_{pk} / \Omega_{beam}) \lambda^2 / 2k = 1.22\times10^3  S_{\nu} (mJy/beam) / (\theta_{maj} ~\theta_{min} ~ \nu(GHz)^2) = 1.36 S_{\nu} (mJy/beam) ~ \lambda(cm)^2) / (\theta_{maj} ~\theta_{min}) $ where the FWHM beam sizes $\theta$ are in arcsec.  }
\tablenotetext{b}{The brightness temperature of the dust emission contribution at 112.3 GHz after removing the extrapolated synchrotron/free-free seen at longer wavelengths from the observed $T_B$ of 163 K (West) and 39 K (East).}
\end{deluxetable*}

\section{Continuum Emission}

\subsection{Infrared Dust Emission}\label{background}

   A brief background for understanding the far infrared 
dust emission is warranted as a preamble to our analysis below. For this discussion we might visualize the massive dust concentrations in the nuclei 
of Arp 220 as spherical, with density decreasing outwards. The dust 
has an opacity which increases steeply at shorter wavelengths (i.e. $\kappa_{\nu} \propto \nu^{1.8}$ 
in the far infrared/submm regime). The enormous dust column densities in the Arp 220 
nuclear sources imply that the dust will be optically thick well into 
the far infrared. 

As one views these sources from the outside, 
the depth from which the observed, emergent radiation is emitted will depend 
strongly on the wavelength of observation, since at each wavelength one sees into the photosphere at $\tau_{\lambda} \simeq 1$. 
At longer wavelengths the dust 
opacity is less, so the emergent radiation will come from deeper within the 
optically thick cloud. 

The dust at each radius is also likely to be in radiative 
equilibrium with the luminosity (which originates from massive stars in 
the nuclear starburst and/or from a central AGN). As long as these luminosity sources 
are centrally concentrated, the dust temperatures must increase at smaller 
radii. 

Thus it should be anticipated -- as long as the dust is optically thick 
into the far infrared,  the longer wavelength observations will tend to probe 
high temperature dust closer to the central energy sources -- the long wavelengths 
see deeper into the enveloping dust. It would appear counter-intuitive that longer wavelength observations probe hotter regions in the nuclei at smaller radii, but because of the opacity falloff at longer wavelengths,
this is in fact the case.

A cautionary corollary -- lower 
angular resolution observations will tend to include a larger fraction of the 
cooler dust at large radii, and these outer regions will be less optically thick (if the gas and dust 
density decreases outward). In fact, one must recognize that there is no single
wavelength at which one can say the {\emph source} is optically thin or thick -- the dust 
column (and optical depth) increases as the line of sight passes closer to the center;
therefore the 'apparent' (or average) optical depth one infers from an observation 
is dependent on the angular resolution or beam size of the observation.
   
    \begin{figure*}[ht]
\epsscale{1}  
\plotone{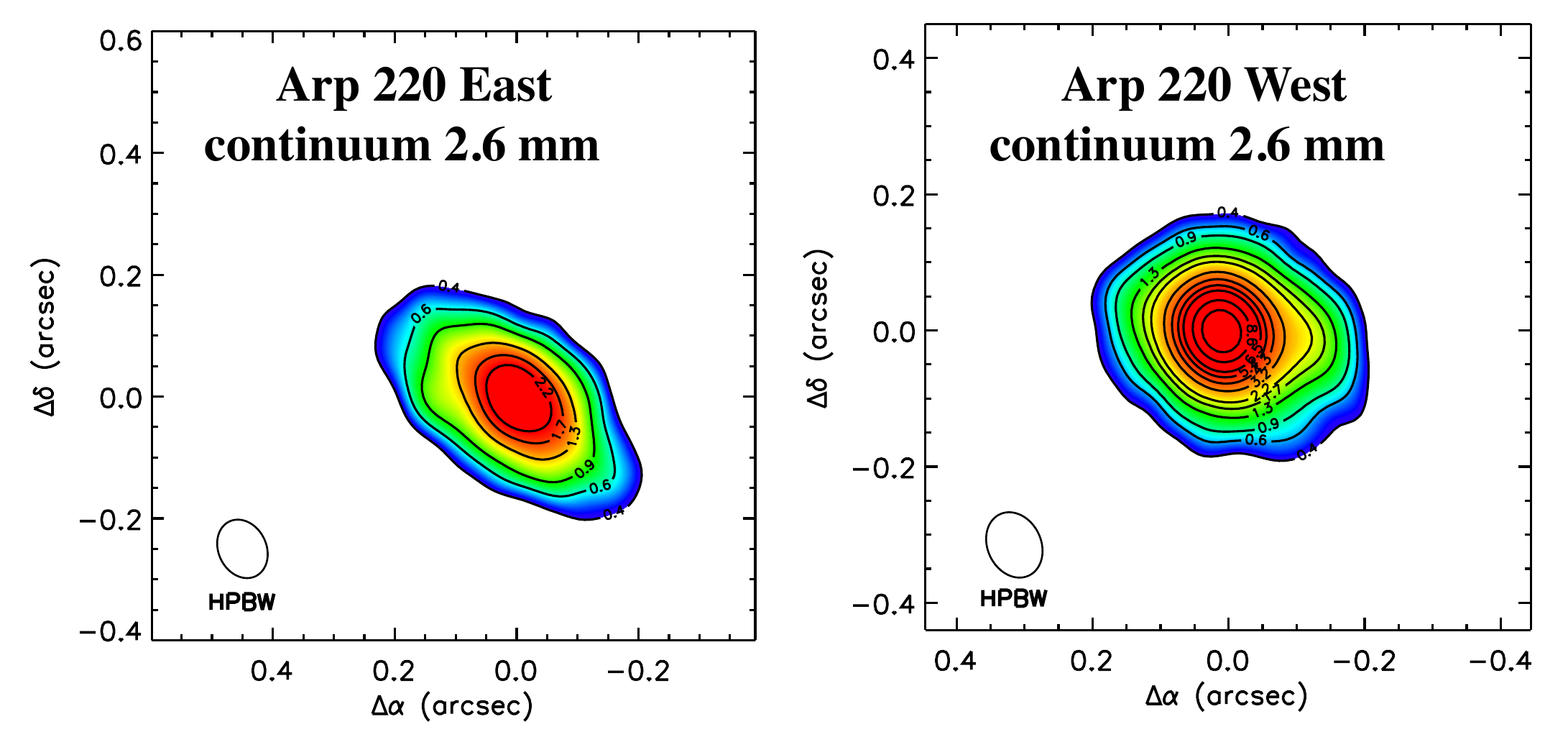}
\caption{The 2.6 mm continuum distribution on the East and West nuclei is shown at $0.08\times0.1$\arcsec resolution ($33\times41$ pc) . The peak values are 3.1 (East) and 13.1 mJy beam$^{-1}$ (West).
Coordinate offsets are relative to the 2.6 mm continuum peaks (Table \ref{fits}) and the contours are mJy beam$^{-1}$.}
\label{cont_image} 
\end{figure*}

\begin{figure}[ht]
\epsscale{1}  
\plotone{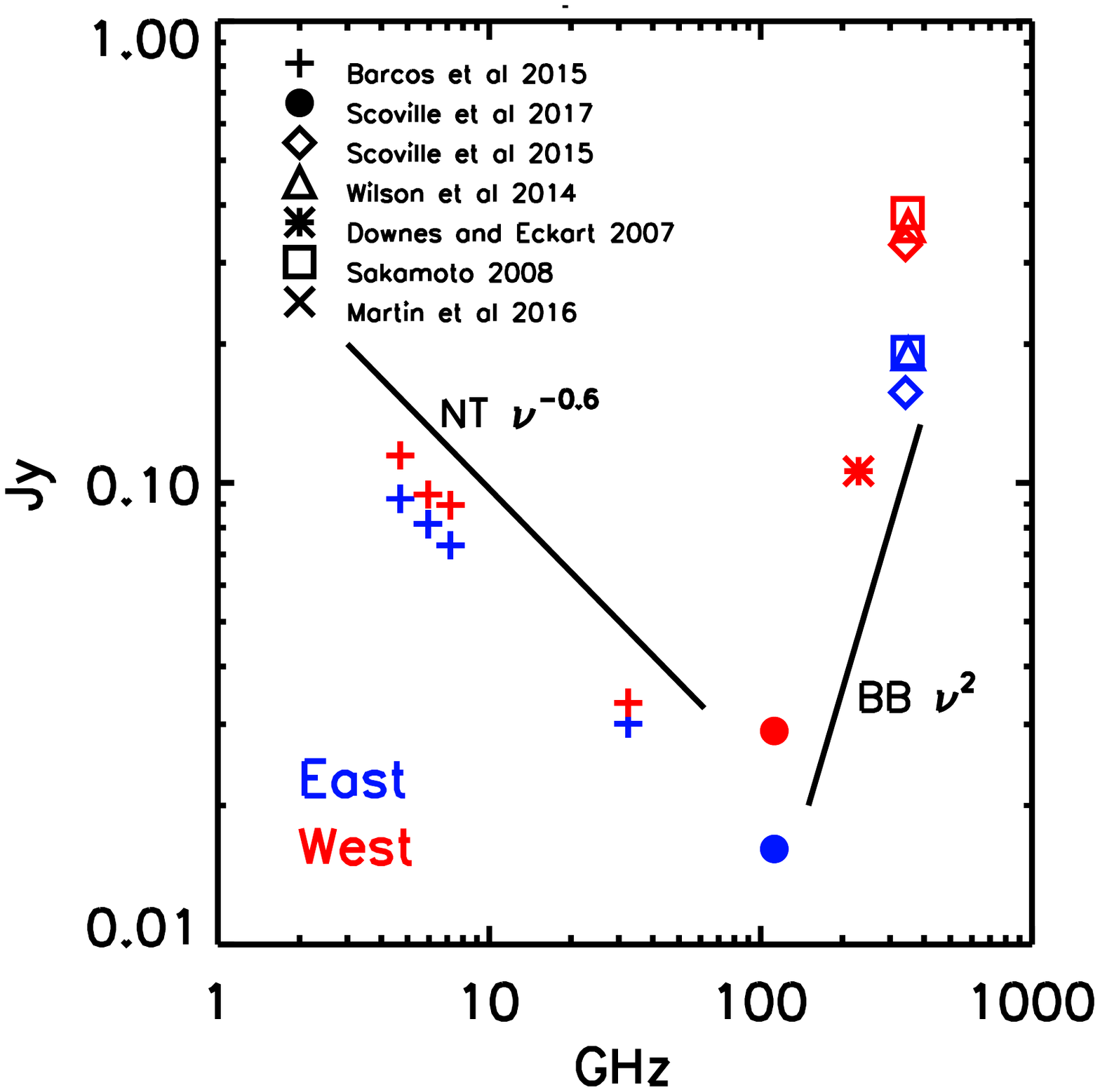}
\caption{The continuum fluxes measured at high resolution for the East and West nuclei of Arp 220 (Table \ref{cont}). 
The expected power laws for non-thermal synchrotron emission (NT) and Rayleigh-Jeans black body (BB) emission are shown for reference. At long wavelengths, 
the dominant emission is non-thermal synchrotron emission with flux varying $\nu^{-0.5 ~\rm to ~-0.7}$; at $\nu > 100$ GHz the dominant 
emission is the dust continuum with spectra varying between $\nu^{2}$ in the optically thick limit and $\nu^{\sim3.8}$ in the optically thin limit. The dust emission at 
${\bf \nu > 100}$ GHz is optically thick in the smallest aperture measurements and a combination of both optically thick 
and thin emission in larger aperture measurements.}
\label{cont_fig} 
\end{figure}

\subsection{Continuum Spectra}

In Table \ref{cont}, we list the radio and infrared continuum measurements done at $<$ 0.6\arcsec resolution -- sufficient to separate 
the two nuclei. The total integrated fluxes and peak flux per beam are given, along with the peak brightness temperature calculated from the peak flux 
per beam at each of the different frequencies. It should be noted that the synthesized beam sizes vary between 0.6 and 0.08\arcsec. In most cases, 
the interferometric integrated fluxes will recover emission on scales up to $\sim 3 \times$ the beam size but emission on larger scales will
not be fully recovered. The brightness temperatures determined from the peak flux at each frequency also refer to variable 
source radii ranging from 0.05 to 0.3\arcsec, corresponding to 20 to 123 pc. 

%{eqnarray}
%T_{B} & =& ~1.36 \left({\lambda (cm) \over{ \theta (\arcsec) }}\right)^2 S_{\nu} (mJy/beam) .
%\label{lnu_eqn}
%\end{eqnarray}   

In Figure \ref{cont_fig} the fluxes are plotted and the spectral indexes for each frequency interval are shown. At low frequencies ($\nu <$ 40 GHz) the spectral index ($\alpha$ in $ S_\nu \propto \nu^\alpha$)
is negative and the emission is predominantly non-thermal synchrotron. At $\nu \geq 200$ GHz the dust emission dominates and the spectral indexes ranging 
from $\alpha = 2$ (optically thick and a single dust temperature) to steeper ($\alpha > 2 ~\rm to \lesssim 4$) if there are substantial contributions from optically thin dust. 
In this frequency range the maximum spectral index will be $\sim 3.8$, corresponding to optically thin emission with a typical Galactic ISM dust opacity coefficient $\beta = 1.8 \pm 0.1$ \citep{pla11b}. In Figure \ref{cont_fig}-right, the spectral indexes at the highest frequency are 2.4 (West) and 3.2 (East), indicating 
that most of the emission is from optically thick dust with a small addition of optically thin emission (most obvious in the East nucleus where the optical depths are 
not so high). 

\subsection{Dust Emission Fluxes and Brightness Temperatures}

At 112.3 GHz, the continuum emission is a mixture of synchrotron and dust emission. If we extrapolate the 32 GHz flux measurements \citep{bar15} with a spectral index $\alpha = -0.60$ determined 
between 7 and 32 GHz (see Figure \ref{cont_fig}), then the expected non-thermal and free-free contribution at 112.3 GHz will 
be $\sim$0.47 of the 32.5 GHz fluxes, implying 15.8 mJy and 3.1 mJy beam$^{-1}$  in the West nucleus and 14 mJy and 1.9 mJy beam$^{-1}$ in the East nucleus. Subtracting these contributions from the observed fluxes, we obtain estimates for the dust continuum fluxes of  13 mJy and  10 mJy beam$^{-1}$ (West) and 1.8 mJy and 1.2 mJy beam$^{-1}$ (East), at 112.3 GHz. These dust-only fluxes are used to calculate the revised dust emission brightness temperatures given in parenthesis in Table \ref{cont} for 112.3 GHz -- 121 K in the West nucleus and 15 K in the East on the scale 90 mas diameter or 19 pc radius.  The peak brightness temperatures in the CO (1-0) 
emission are 187 and 175 K, respectively (see Table \ref{fits});  these peak line brightness temperatures are seen at larger radii (see Figure \ref{spec_fig}).

\subsection{West Nucleus Luminosity}\label{west_dust}

The 147 K brightness temperature of the 112.3 GHz dust emission seen on the West nucleus is unexpectedly high, given the previous submm observations (691 GHz) of \cite{wil14} which indicated 
T$_B = 181 \pm 27$ K on the West nucleus and optical depth $\tau_{434\mu m} \simeq$ 5.3. If the dust opacity varies with the standard Galactic power-law $\nu^{1.8}$, then the 
optical depth should be $\sim 0.038 \times 5.3 \sim 0.2$ at 112.6 GHz. The estimated 147 K brightness temperature clearly requires a higher optical depth $\tau_{2.6 mm} \simeq 1$. 

It is interesting to note that the implied dust photosphere, optically thick out to $\lambda \sim 2.6$mm, with dust temperature $T_D \simeq 200$ K and radius $\simeq 15 $pc, must 
radiate 

\begin{eqnarray}\label{lum_eqn}
L_{\rm West ~200K ~dust} & =& ~4\pi R^2 \sigma T^4  \nonumber \\
 & =&  6.35\times 10^{11} \lsun \times   \nonumber \\
 & & (R/15~pc)^2 (T/200~K)^4 .
\end{eqnarray}

\noindent  An identical estimate, L$_{\rm West} = 6.3\times10^{11}$ \lsun, was obtained by \cite{wil14} from ALMA dust continuum measurements at $\lambda = 445 \mu$m.  This corresponds to 33\% of the total IR luminosity $1.91\times10^{12}$ \lsun ~of Arp 220 \citep{san03,arm09} -- all originating from R $\leq $15 pc. 

 Based on their 860 $\mu$m continuum measurements at 0.23\arcsec~ resolution, \cite{sak08} obtain lower limits on the luminosity $\sim3\times10^{11}$ \lsun~ for the West nucleus. 
They argue that the derived luminosity is very sensitive to the adopted source size, varying as R$^{-6}$. Our estimate given above is based on a resolved source size 
and a dust temperature $\sim200$ K, based on determination that the dust is nearly optically thick. It therefore is not just a lower limit but rather an estimate of the 
actual luminosity, albeit uncertain to a factor 2.    

For a geometrically thin, optically thick disk, the emitting surface area is $\sim2\pi R^2$, so for the same $T_D$ the emergent luminosity would 
be a factor 2 lower than Equation \ref{lum_eqn}. It is interesting to note that essentially all estimates of the infrared luminosity in ULIRGs assume 
a spherical, i.e. isotropic source. If the infrared is in fact emitted from a thin disk which is optically thick, these luminosity estimates should be increased by a factor $\rm 1/cos~ i$.

A considerably lower  temperature $T = 45$ K is derived from fitting the 
total far infrared SED of Arp 220 \citep{san91}, implying that, at the shorter wavelengths near the 70 $\mu$m IR peak, one is sampling colder dust in a more extended 
photosphere. This sampling of colder dust at shorter wavelengths is as anticipated in Section \ref{background} for an optically thick dust cloud --  the optical depth is higher at shorter wavelengths. Hence, the $\tau \simeq 1$ surface (from 
which the observed photons at shorter wavelengths originate) will be at larger radius 
and lower T (assuming a centrally heated source). 

A major virtue of the longer wavelength observations reported here is the ability to penetrate the optically 
thick dust envelope and probe the inner regions of heavily obscured luminosity sources. The peak of the 200 K blackbody emission ($L_{\nu}$) is at $\lambda = 26\mu$m but the original 
sources of luminosity (young stars and/or a central AGN) undoubtedly emit their energy at much shorter wavelengths in the visible, UV and X-rays. 

Since the observations place an upper limit of $\sim200$ K on the dust temperature at R = 15 pc, this implies that the luminosity originating from smaller radii cannot
be much larger than $6\times10^{11}$ \lsun -- otherwise the dust would be hotter. If there is an AGN in the Western nucleus its power must be less than this. The remainder of the $\sim1.9\times10^{12}$\lsun ~total luminosity, or about $1.3\times10^{12}$ \lsun, must originate from more distributed star formation at R $> 15$ pc (36 mas radius) {\bf and the East nucleus}.

\subsection{West Nucleus ISM column densities and Mass}\label{west}

We use the observed dust opacity ($\tau_{2.6 mm} \simeq 1$) to estimate the column density of the West nucleus point source
assuming that the dust there has similar properties to general ISM dust observed in the Galaxy.  \cite{sco16} empirically calibrated the 
long wavelength dust emission from Herschel SPIRE with the CO(1-0) ISM masses for 28 local star forming galaxies and 12 ULIRG galaxies and with Planck measurements of the dust submm dust opacity in the Milky Way. The data are consistent with a single proportionality constant relating the rest-frame 850$\mu$m specific luminosity of the dust to the molecular ISM mass. This empirical calibration is 

\begin{eqnarray}\label{alpha}
\alpha_{\nu} &\equiv& < L_{\nu_{850\mu \rm m}} /M_{\rm mol}>    \nonumber \\
 &=& (6.7\pm1.7)\times 10^{19} \rm erg ~sec^{-1} Hz^{-1} {\msun}^{-1} 
\end{eqnarray}

\noindent where $L_{\nu_{850\mu \rm m}}$ is the specific luminosity of the dust at $\lambda = 850 \mu$m. The 
mass $M_{mol}$ includes contribution for He and heavier atoms and the mass derivation employed a single standard Galactic CO-H$_2$ conversion factor. 

For an optically thin mass sheet of area A and surface density $\Sigma_{mol}$  ISM, Equation \ref{alpha} can be recast as 
\begin{eqnarray}\label{alpha1}
 L_{\nu_{850\mu \rm m}} /M_{\rm mol} &=& {2 \pi A B_{\nu} (T_D) \tau_{\nu} \over{ A \Sigma_{mol}}}   \nonumber \\
 &=& {2 \pi B_{\nu} (T_D) \kappa_{\nu} \Sigma_{mol} \over{ \Sigma_{mol}}} \nonumber \\
  &=& 2 \pi B_{\nu} (T_D) \kappa_{\nu}  \rm , ~~ yielding \nonumber \\
  \kappa_{\nu}  &=& {\alpha_{\nu} \over {2 \pi B_{\nu} (T_D) }} ~~.
\end{eqnarray}

We use the above calibration of dust optical depth to estimate the column of gas in the nuclear source. The bulk of the ISM dust in the nearby galaxies 
used for the calibration is at $\sim$25 K \citep[see][]{sco16}. Using this temperature in the above equations then implies a dust absorption coefficient at 850 $\mu$m $\kappa_{850\mu m} = 8.06
\times10^{-3}$ cm$^2$ gr$^{-1}$ where the mass includes the He contribution. Scaling this opacity coefficient as $\nu^{1.8}$ \citep{pla11b}, we obtain
$\kappa_{2.6mm} = 1.03
\times10^{-3}$ cm$^2$ gr$^{-1}$. Putting this in terms of the H$_2$ column density, $\kappa_{2.6mm} = 4.51\times10^{-27}$ N$_{H_2}$. Thus an H$_2$ column of 
$2.21\times10^{26}$ H$_2$ cm$^{-2}$ is required in order for $\tau_{2.6 mm} \simeq1$.\footnote[1]{Alternatively, the Planck ratio of 
$\tau_{250\mu m} / N_H = 2.32 \times10^{-25}$ cm$^{2}$ derived for Milky Way H$_2$ \citep{pla11a} translates to $1.45\times10^{26}$ cm$^{2}$ H$_2$ cm$^{-2}$ adopting the same $\nu^{1.8}$ dependence of the opacity coefficient.}

 Our estimate of the column density is considerably higher than earlier estimates of N$_{H2} \sim 10^{25}$ and $3 - 6\times10^{25}$ cm$^{-2}$ \citep{sak08,gon12}. 
Their estimates were lower since they were derived from higher frequency observation and hence should be viewed as lower limits if the dust is optically thick at those wavelengths. 
Our estimate is also well above the lower limit of $> 10^{25}$ cm$^{-2}$ derived by \cite{ten15} from non-detections in the NuSTAR bands above 20 keV. 

As an aside, it is interesting to note that the mass column density of ISM $\simeq$900 g cm$^{-2}$ in the Western nucleus corresponds to a concrete wall 3 - 4 m thick or gold 1 ft thick. This is perhaps the highest ISM column density ever probed by astronomical observations. It corresponds to A$_V = 2\times10^5$ mag 
and would be very Compton thick. 

Lastly, we note that $\tau_{2.6mm}$ must be $\simeq1$ across a circular region with R $>$ 10 pc since the similarity of the observed dust brightness temperature and dust physical temperature requires an areal filling factor of order unity in the central resolution element. 
The total molecular mass is then M$_{mol} > 1.4\times10^9$ \msun. Clearly, these estimates are uncertain, given the assumption that the dust 
has standard interstellar dust properties and abundance, relative to gas in the extreme conditions at the center of Arp 220. The mass so derived is 
approximately a factor 2 higher than the dynamical mass in the same region (also uncertain) obtained from the CO line kinematics (see  Section \ref{modeling}).  

\begin{figure*}[ht]
\epsscale{1}  
\plotone{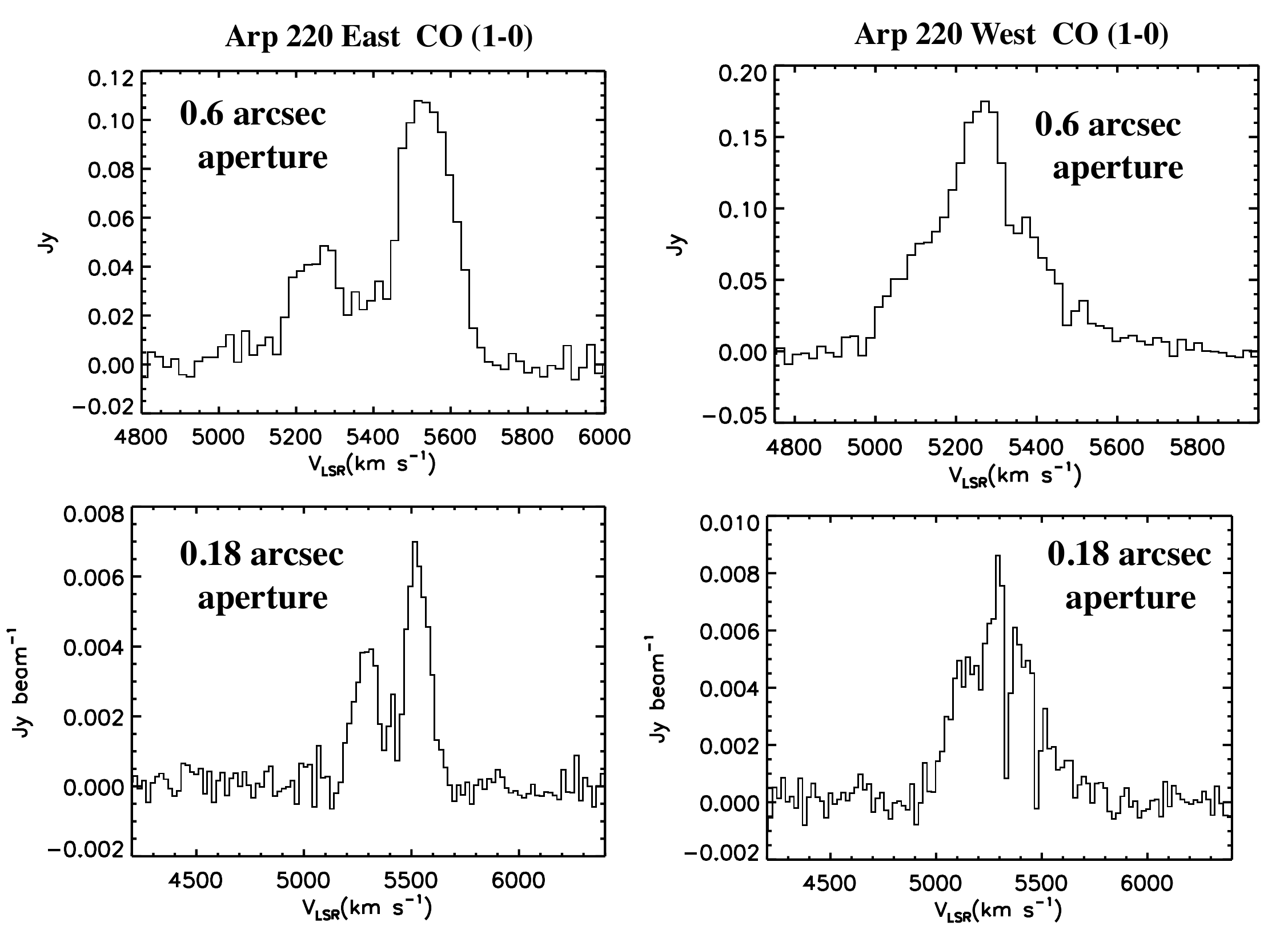}
\caption{The CO (1-0) emission from the East and West nuclei, in apertures of 0.8 \arcsec (top panels) and 0.18 \arcsec (bottom panels) diameter centered on each nucleus. The emission extends over 600 to 800 \kms in both nuclei. Narrow self-absorption 
features can also be seen in some spectra (e.g. lower right panel). For the top panels, the flux in Jy is the integrated flux within the aperture; for the lower
panels where the aperture is comparable with the beam size, we plot the average pixel value within the aperture (i.e. Jy beam$^{-1}$). Note that in the bottom spectra the velocity range is much larger.}
\label{spec_fig} 
\end{figure*}

From the derived H$_2$ column density N$_{H_2} = 2.21\times10^{26}$ $H_2$ cm$^{-2}$ and line of sight path length of 30 pc, we infer a mean volume 
density n$_{H_2} = 2.4\times10^6$ cm$^{-3}$
in the central area of the West nucleus. 

At these extraordinarily high ISM densities, the dust and gas will be collisionally coupled 
(at $n_{H_2} > 10^4$ cm$^{-3}$) and in thermal 
equilibrium ($T_D = T_k$).  Most molecular transitions at mm/submm wavelengths will have level populations in thermal equilibrium with the $H_2$ gas. And if the dust is optically thick into the mm regime, there will be substantial direct radiative coupling 
of the mm transitions to the dust radiation field. This is all consistent with the 
observed 187 K CO brightness temperature (see Table \ref{fits}). 

%An alternative approach to estimate the mass from the dust emission would be to use the calibration between dust Rayleigh-Jeans flux and mass (Equation \ref{alpha})
%and to scale the emissivity by a factor $\sim200$ K/ 25 K (to account for the higher dust temperatures in the compact nuclear regions compared the bulk 
%galactic ISMs used for the calibration \citep{sco16}). The route used above -- working directly from the optical depth -- avoids this uncertainty in scaling for  variations in the elevated dust temperatures. 

\subsection{The East Nucleus}

 \begin{figure*}[ht]
\epsscale{1}  
\plotone{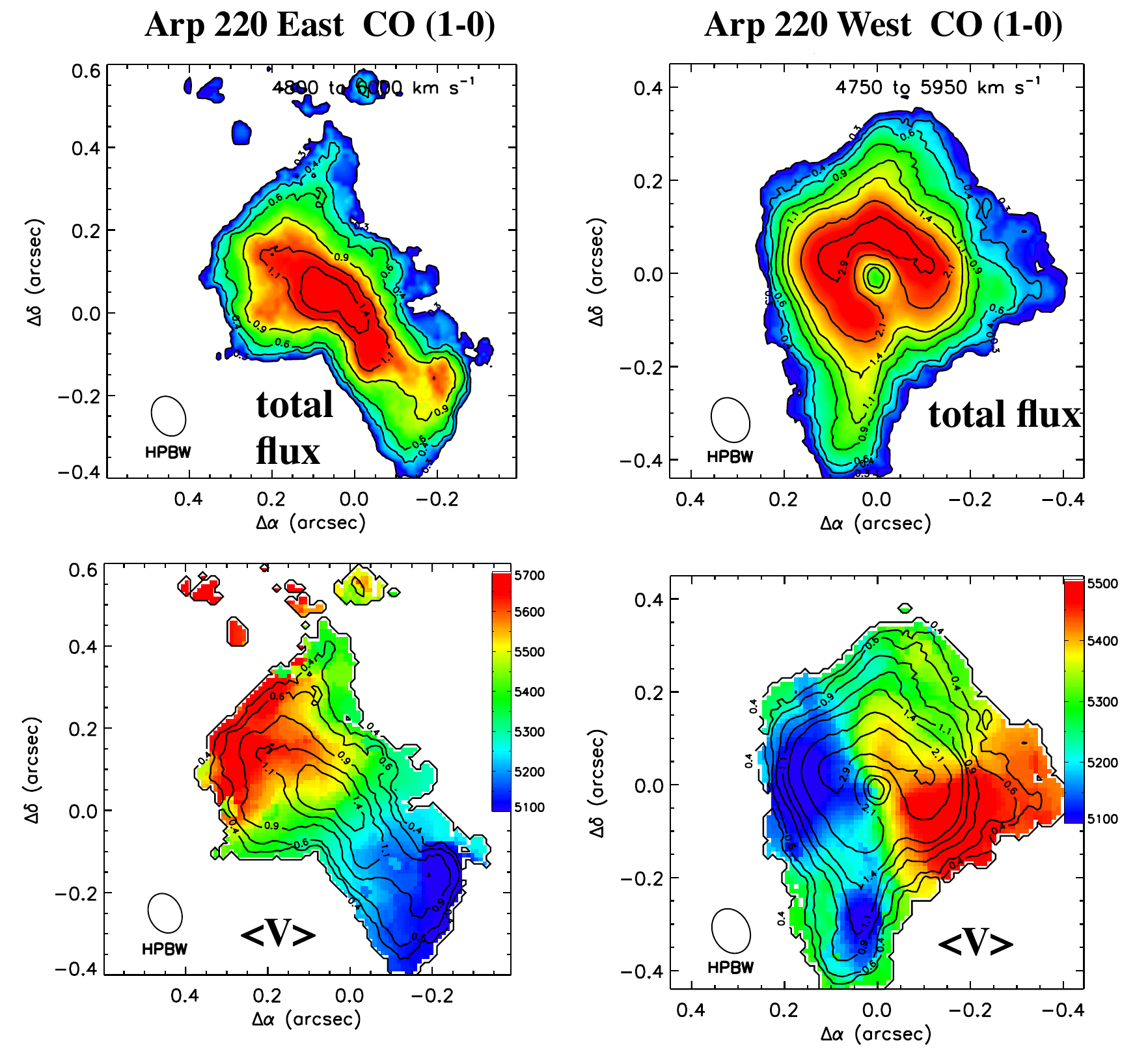}
\caption{The integrated CO (1-0) line flux (top panels) and mean velocity (bottom panels) are shown.
In the East nucleus the emission is clearly elongated along the major axis of the kinematics. In the West 
nucleus a hole is seen in the center coinciding with the dust continuum peak {\bf (see Section \ref{hole})}. 
These images were 
computed with a clipping cut to exclude from the line integrals any pixels below the 4$\sigma$ noise level.
The coordinate offsets are measured relative to the 2.6 mm continuum peaks (see Figure \ref{cont_image}).
The contours are labelled with Jy beam$^{-1}$ and \kms. }
\label{line_fig} 
\end{figure*}

For the East nucleus the dust optical depth is less. \cite{wil14} estimate $\tau_{434\mu m} =  1.7$ and T = 80 K (compared to the above-mentioned values of 5.6 and 181 K for the West nucleus). 
This is borne out in our 2.6 mm continuum imaging which indicates a peak of just 15 K for the dust in the East nucleus (after removing the synchrotron and free-free contributions). This suggests $\tau_{2.6 mm} \sim 0.2$ (assuming $T_D = 80$ K). In this nucleus, we can't provide the equivalent energetic constraints as in the West from our data since the dust is not optically thick 
at $\lambda = 2.6$ mm. However, from the ratio of $\tau$s derived by \cite{wil14}, one might infer that the East nucleus has $\sim 1/4 $ the luminosity of the West nucleus. 
Similarly, the ratio of dust opacities in the two nuclei suggests that
the mass of dust and gas in the East nucleus is $\sim 1/4$ of that in the West nucleus. Better constraints will 
be provided by the high resolution CO(2-1) imaging with ALMA, which is scheduled.

\section{CO (1-0) Line Emission}

Figure \ref{spec_fig} shows the continuum-subtracted CO (1-0) emission line profiles for the West and East nuclei obtained in apertures of 0.8 \arcsec
diameter centered on each nucleus. The CO emission extends up to $\sim800$ \kms in each of the nuclei and is offset in the mean by $\sim 120$ \kms ~between the two nuclei. In the West nucleus the line profile exhibits broad wings and a single 
peak while in the East nucleus it is double peaked.

Images of integrated CO emission and the intensity weighted centroid velocity ($\langle V \rangle$) are shown in Figure \ref{line_fig}.  The overall morphology 
of the gas distribution is remarkably different in the two nuclei -- in the West nucleus, the total emission is less elongated along a major axis; there is a drop in the 
CO emission on the central resolution element (90 mas diameter); and overall, the emission is more compact (as was the case for the dust continuum). 

Despite these differences, both nuclei exhibit a clear kinematic gradient, suggesting rotation. In the East nucleus, the kinematic gradient aligns closely 
with the major axis of the elongated CO intensity distribution. In the West nucleus the kinematic major axis is at PA $\simeq -110 \deg$; in this 
nuclear source the emission intensity distribution is hardly elongated and one does not see correlation with the major kinematic axis of either the CO 
or the continuum (see Table \ref{fits}). The magnitude of the mean velocity gradients is impressive;  in both nuclei, the shift is $\sim500 - 600$ \kms (see Figure \ref{line_fig}-lower panels) over 0.3 -- 0.4 \arcsec (124 -- 165 pc).

\subsection{CO Hole on the West Nucleus}\label{hole}

In the West nucleus, the central dip in the CO emission is coincident with the central dust continuum peak. The depth of the 
hole is approximately a factor three compared with the immediately exterior ring (see Figure \ref{line_fig}-upper right). Since the 
dust continuum in the West peaks strongly in the central resolution element, this dip in the CO emission is probably not due to 
a deficiency or clearing of ISM at small radii. 

There are three possible explanations for the drop in the strength of the CO emission at the center: 1) the CO and the dust are in thermal equilibrium in this high density 
core; 2) the high excitation temperature of the CO in the core depletes the lowest CO rotational levels, in favor of much higher J states causing the CO (1-0) line to be optically thin; 
and 3) the CO emission from the core is self-absorbed by colder CO in the foreground along the line of sight. 

The first explanation is consistent with the very high column 
density, $\rm N_{H2} \simeq 2\times10^{26}$ H$_2$ cm$^{-2}$  and volume density n$_{H2} = 2.4\times10^6$ cm$^{-3}$ deduced for the dust emission in Section \ref{west}.
At these densities the H$_2$ will be collisionally coupled to the grain temperature and $T_K = T_D$ and the CO levels will be thermalized, i.e. $\rm T_x = T_K$. In this case, 
no line emission will appear in excess of the optically thick dust emission. This does require that the density must drop steeply at the outer radius of the dust photosphere
to avoid there being an external chromosphere producing excess CO line emission outside the photosphere. 

The second explanation, that the CO (1-0) line in the core has lower optical depth than the dust emission, could also follow from the very high excitation conditions 
in the core, compared to those in normal Galactic molecular clouds. This, combined with the large velocity widths expected in the core, could cause the CO (1-0) to be optically thinner than the dust at $\lambda = 2.6$ mm. In low density Galactic GMCs the CO (1-0) line typically has more than factor $10^{4}$ greater optical depth than the dust at $\lambda$ = 2.6 mm. In the Arp 220 nuclei this ratio is reduced by two orders of magnitude due to the $\sim100$ times higher gas velocity dispersion and another factor of 20 - 40  
due to the $\sim20 - 40$ times higher gas temperature which spreads the CO molecules over more levels. In addition, the stimulated emission correction to the optical 
depth will cause a further reduction. A modest depletion in the CO abundance could then cause the 1-0 line to be optically thin. 

The third explanation, having the CO emission from the core be self-absorbed by low excitation CO further out along the line-of-sight, requires that the foreground gas 
be coherent with that in the core, i.e. at the same radial velocity. The CO emission at small radii will be predominantly at high velocities and thus is not coherent with gas close to the 
systemic velocity outside the nucleus. 

In Arp 220 East, we do see three narrow absorption features (spatially offset 0.1 to 0.2\arcsec from the nucleus)  
within $\pm100$ \kms of the systemic velocity; these absorptions are sharp in velocity and never cover more than $\sim30$ \kms (e.g. Figure \ref{spec_fig}-lower left).
There is negative velocity absorption seen in CO (3-2), SiO (6-5) and HCO$^+$ (3-2) which is interpreted as an outflowing wind \citep{sak09,gon12,tun15,ran15,aal15}. However, in order that the wind produce the 
apparent hole in CO (1-0) only from the nucleus, the wind must be confined to radii $\lesssim 16$ pc. This size is inconsistent with the larger spatial 
extent of the high J absorption lines. In addition, the outflow gas would only absorb the nuclear CO emission at negative velocities, leaving the other half of the emission unabsorbed. 

Lastly, we note that if there were CO (1-0) absorption {\bf of the background continuum}, we should expect that in the \emph{continuum-subtracted} 1-0 spectra, we should see 
instances where the line emission has an apparent negative intensity (since too much continuum would have been subtracted at the line velocities). This is not seen 
in the spectra shown in Figure \ref{spec_fig} or \ref{west_spectra}. {\bf If the foreground CO was absorbing only the background line emission from the nucleus, the continuum-subtracted spectra 
would not necessarily go below zero, but it is not clear why the foreground absorption 
wouldn't also be absorbing the nuclear continuum. The 'negative' intensity CO line 
could be suppressed if the continuum and CO line emission had different spatial extents. }

We thus favor the first or second explanations to reasonably account for the CO hole in the West nucleus.The third explanation appears inconsistent 
with the high spatial resolution spectra, requires overlap of radial velocities between the nucleus and the foreground gas and if the absorption takes place in an outflowing wind, then the positive radial velocity emission would be left unabsorbed. 

% in the CO and   
%Instead, most of this drop in the CO emission is likely due to dust absorption of the line emission originating inside the central R $\simeq 15$ pc (see Section \ref{west_dust}). However, in order for the CO emission to decrease in the center relative to the dust emission, 
%the CO must also have a lower optical depth per unit column density of ISM than the dust at $\lambda = 2.6 $mm. Having the dust be optically thick in the center is not \emph{sufficient} to produce the central dip; the CO line must be less optically thick than the dust. 

%These approximate considerations suggest that the CO (1-0) line 
%will not have a lower optical depth than the dust per unit column of ISM, unless the gas phase abundance of CO relative to dust is reduced by at least a factor 5 -- 10 from the standard GMC values. %Although these effects tend in the right direction, they may 
%be insufficient, requiring that the gas phase abundance of CO be reduced also. 

%In proto-planetary disks the CO abundance is very reduced 
%due to depletion on dust grains; however this effect is limited to much lower dust temperatures ($T_D < 30$ K) than are relevant to the West nucleus
%\citep[e.g. see][]{bec90,reb15}. A full analysis of the CO abundance variations in Arp 220 will be possible when the scheduled ALMA observations
%for Bands 6 and 7 are obtained in our program. These will include CO, $^{13}$CO and C$^{18}$O (2-1) and (3-2). 

\subsection{CO Elongation in the East Nucleus}

In the East nucleus the CO emission is clearly elongated with a major/minor axis ratio $\sim$3:1 (see Figure \ref{line_fig}-upper left)
and 2:1 from the Gaussian fitting (see Table \ref{sizes}). If the structure is interpreted as an inclined disk, the former axis ratio
implies an inclination of 70\deg to the line of sight. The East nucleus structure is clearly not axisymmetric -- the peak 
in the CO emission is displaced $\sim0.1$\arcsec NE of the dust continuum peak (coordinate $\Delta \alpha, \Delta \delta = 0,0$ in Figure  \ref{line_fig}). 
There is also $\sim50$\% more integrated CO emission luminosity on the NE side than on the SW.

\begin{deluxetable}{lllll}
\tabletypesize{\scriptsize}
%\normaltsize
%\rotate
\tablecaption{\bf{\underline{Nuclear Disk Emissivity and Kinematic Models}}\label{models}  }
\tablewidth{0pt}
%\tablehead{  \colhead{Parameter} &} \\
%\tablehead{  \colhead{$\nu_{obs}$} & \colhead{Beam} & \colhead{total Flux}  & \colhead{peak Flux}  & \colhead{peak $T_B$\tablenotemark{a}}  & \colhead{reference}  \\
%  \colhead{GHz} &  \colhead{$\arcsec \times \arcsec$}  & \colhead{mJy} &  \colhead{mJy/beam} & \colhead{K} }
\startdata

\\
&  \multicolumn{2}{l}{\bf{\underline{West Nucleus}}}\\
 \\
 Systemic velocity &  V$_{sys} $ & 5337 \kms \\
Gas turbulence & FWHM $\Delta v$ & 250 \kms \\
Disk inclination &  i & 30\deg  \\
Major axis & P.A. & 265\deg \\
Rotation curve: & Point mass & $\sim8\times10^8$ \msun \\
 & Mass at $<$ 70 pc & $\sim1.5\times10^9\msun$  \\
 CO emissivity: 
&  Peak & at R$ < 5$ pc \\
 & Flat and 10$\times$ lower  & at  R = 10 - 50 pc \\
 & axisymmetric. & \\
\\
\\
&  \multicolumn{2}{l}{\bf{\underline{East Nucleus}}}\\
 \\
Systemic velocity &  V$_{sys} $  &  5431 \kms \\
Gas turbulence & FWHM $\Delta v$ & 120 \kms \\
Disk inclination &  i & 45\deg  \\
Major axis & P.A. & 50\deg \\
Rotation curve: & Point mass & $< 10^8$ \msun \\
& Mass at $<$ 90 pc &  $1.5\times10^9\msun$ \\
 CO emissivity: & 0 & at R$ < 5$ pc\\
&  \multicolumn{1}{l}{  Peak }& at R$ \sim$ 10 pc \\
&  \multicolumn{1}{l}{  Falls a factor 2  } & out to 100 pc\\
&  \multicolumn{2}{l}{ Receding side $2\times$ }\\
&  \multicolumn{2}{l}{  brighter.}\\
  \enddata
%\tablecomments{}
%\tablenotetext{a}{}
%\tablenotetext{b}{}
\end{deluxetable}

\subsection{Major Axis Kinematics}

\begin{figure*}[ht]
\epsscale{1}  
\plotone{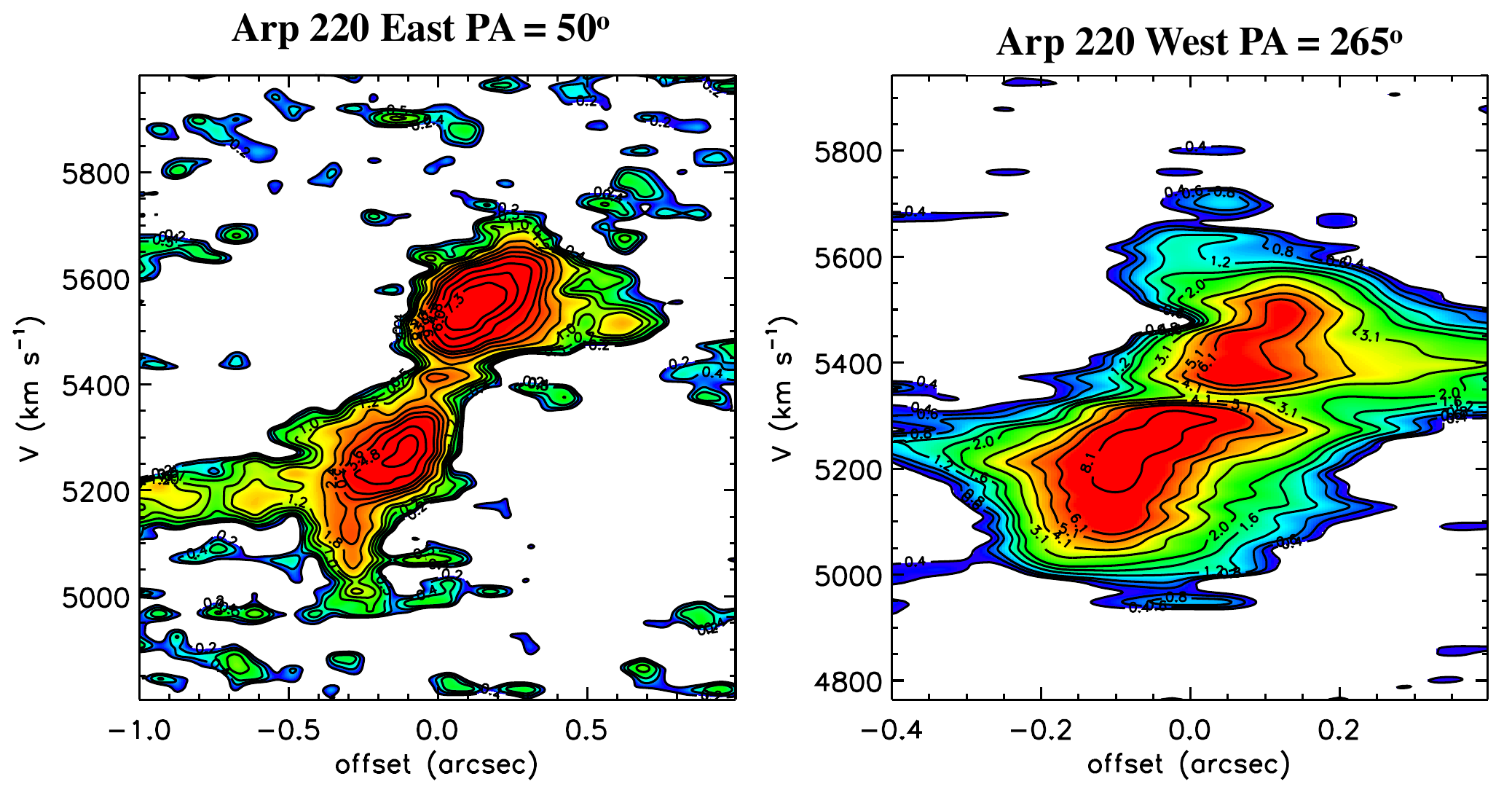}
\caption{Spatial-velocity strip maps along the major axes of Arp 220 East (PA=50\degg) and West (PA = 265\degg). 
The coordinate offsets are measured relative to the 2.6 mm continuum peaks (see Figure \ref{cont_image}). Positive 
offset coordinate corresponds to the receding (redshifted) side on the major axis; i.e. for the West nucleus, positive offset is to the west
and for the East nucleus, positive offset is northeast. Thus, on the West strip map (right panel), the emission at offset = -0.4\arcsec and 5250 \kms 
is emission from the East nucleus coming in, and on the East strip map (left panel), the emission at offset $<$ -0.5\arcsec and 5200 \kms is the 
West nucleus coming in. %(NOTE -- the right figure has a funny offset between the contours and the color image which i need to track down.)
The contours are labelled with mJy beam$^{-1}$. }
\label{rot_curve} 
\end{figure*}

Figure \ref{rot_curve} shows the distribution of CO emission and the gas kinematics along the major axes of the 
West and East nuclei. The central reference position is taken to be the 2.6 mm continuum peak in each nucleus. 
In both nuclei the velocity gradient extends over approximately 700 \kms. In the West nucleus, this velocity range is
seen within $\pm0.1$\arcsec (R = 41 pc) of the center; in the East, it occurs within $\pm0.4$\arcsec (R = 165 pc) of the center.
Figure \ref{spec_fig} illustrates the contrast in spatial extent of the high velocity emission between the two nuclei comparing the top and bottom panels.
The strip maps also clearly show the decrease in line emission on the nuclear peaks at the central velocities. At the same time, 
the central positions show a broad range of emission velocities: in the West, 4950 to 5700 \kms ~and in the East,  
5150 to 5700 \kms. This is also seen in the mean spectra for the nuclei shown in Figure \ref{spec_fig}-lower panels. 

\section{CO Emission Distribution and Kinematics}\label{modeling}

 \begin{figure*}[ht]
\epsscale{1}  
\plotone{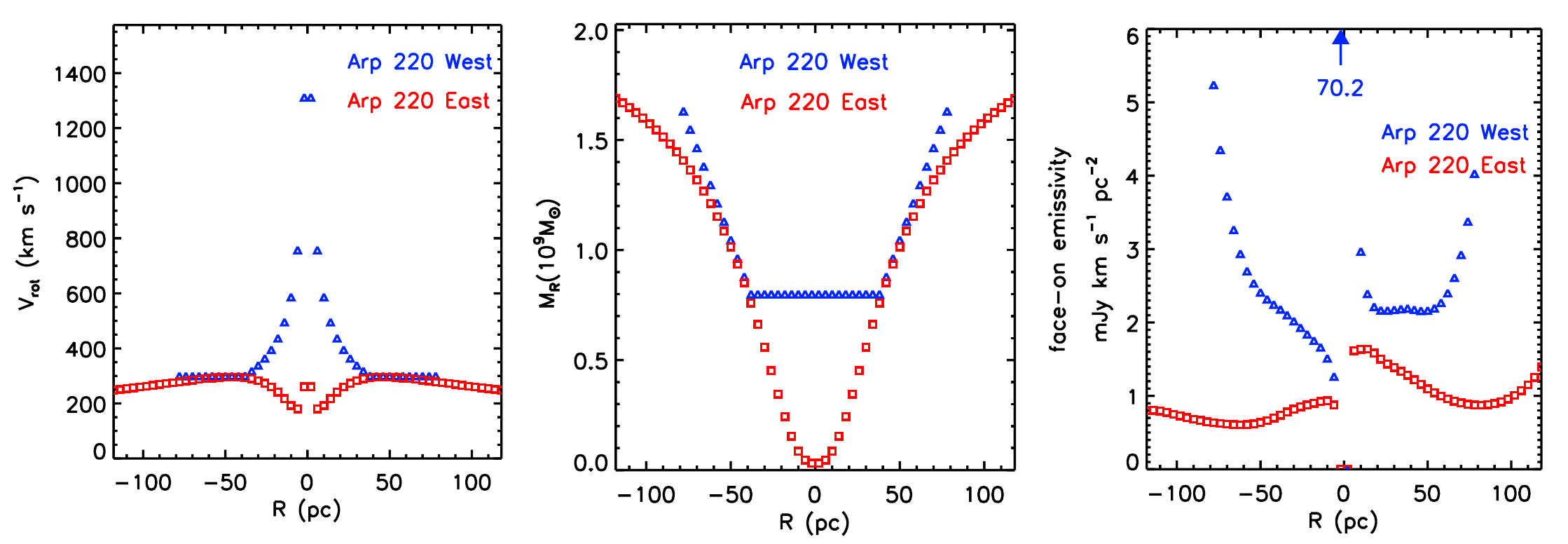}
\caption{The derived rotation curves, enclosed dynamical mass as a function of
radius are shown the Arp 220 East and West disks as derived from fitting the high resolution line profiles (see text). The dynamical mass is obtained in the spherical approximation with $M_R = RV_{rot}^2/G$. The emissivity distribution (per pc$^2$ of area) is that which would be seen if the disks were face-on to the line of sight. The observed total-integrated-flux is the integral of this 
face-on emissivity corrected for inclination (i.e. multiplied by cos i).
A central point mass of $8\times10^8$ \msun~ is required in the West nucleus and only a limit of $< 10^8$ \msun ~in the East nucleus. The CO (1-0) emissivity shown in the right panel is fairly flat with radius, except for the high emissivity central point required in the West nucleus.}
\label{rot_curve1} 
\end{figure*}

In order to understand the small scale distribution of the CO emission and to place constraints on the gas kinematics, we have modeled 
each of the two nuclei with rotating disks inclined to the line of sight. This model, including a parametrized rotation curve and 
gas velocity dispersion, was then fit to the observed CO line profiles on a well-sampled grid (60 mas spacing) in each nucleus, in order to derive the
best fit radial distribution of CO line emissivity. This was carried out using the maximum likelihood procedure developed in \cite{sco83}
with the modification that emissivities were allowed to be different on the + and - offset sides of the major axis. 

 This modeling is with a 2D, infinitesimally thin planar disk with 
constant velocity dispersion as a function of radius. This is obviously a much simplified model to describe the central disks in a merger system with a 
nearby companion and with energetic feedback from SF and active nuclei feedback. The latter could be expected to result in velocity dispersion increasing 
toward the center, and the former would certainly warp any planar disk structure. However, given the limited resolution of the present observations relative to 
the scale of the disk, such higher order modeling is not yet warranted.

 The models developed 
for the high resolution CO (1-0) emission provide acceptable fits to the collection of observed CO (1-0) line profiles with an overall reduced $\chi^2 \sim4-6$ for both disks. The observed and model spectra are shown in Appendix \ref{co_spectra}. This  
is a good fit, given the simplicity of the model (bi-axisymmetric with a single velocity dispersion and rotation curve). The  preliminary modeling results  are shown in Figure \ref{rot_curve1} and summarized in Table \ref{models}. They establish the context for the discussion of the physical structure of the disks in Section \ref{physics}.

 The dynamical mass estimates were calculated assuming a spherical configuration (see Section \ref{dyn_mass}). The noteworthy constraints from the disk line profile modeling are: 1) a central point mass $\sim8\times10^8$\msun ~is apparently required in the 
West nucleus to provide a best fit to the high velocities observed in the center, but no central point mass is required for the East nucleus ($< 10^8$ \msun) and 2) 
the overall dynamical masses are $\sim1.5\times10^9$\msun ~at radii less than 70 and 90 pc, respectively. We expect that these parameters 
may change somewhat when we simultaneously model the other CO transitions and include the lower resolution imaging, but they set a framework for 
present models. The full modeling with the complete datasets will be presented in \cite{sco17}.

 Two cautions are in order with respect to the possible existence of a point mass in the West nucleus. First, our modeling has assumed a constant velocity dispersion 
in each disk. This was done out of necessity to simplify the modeling. A large increase in the velocity dispersion (due to feedback) at the nucleus, cannot 
be ruled out as a source of the large velocities at small radius. Secondly, the apparent point mass might be the result of extreme gas settling into the nucleus, i.e. 
the large mass could be an interstellar gas concentration and not necessarily a black hole.

\subsection{Dynamical Masses}\label{dyn_mass}

 \begin{figure}[ht]
\epsscale{1}  
\plotone{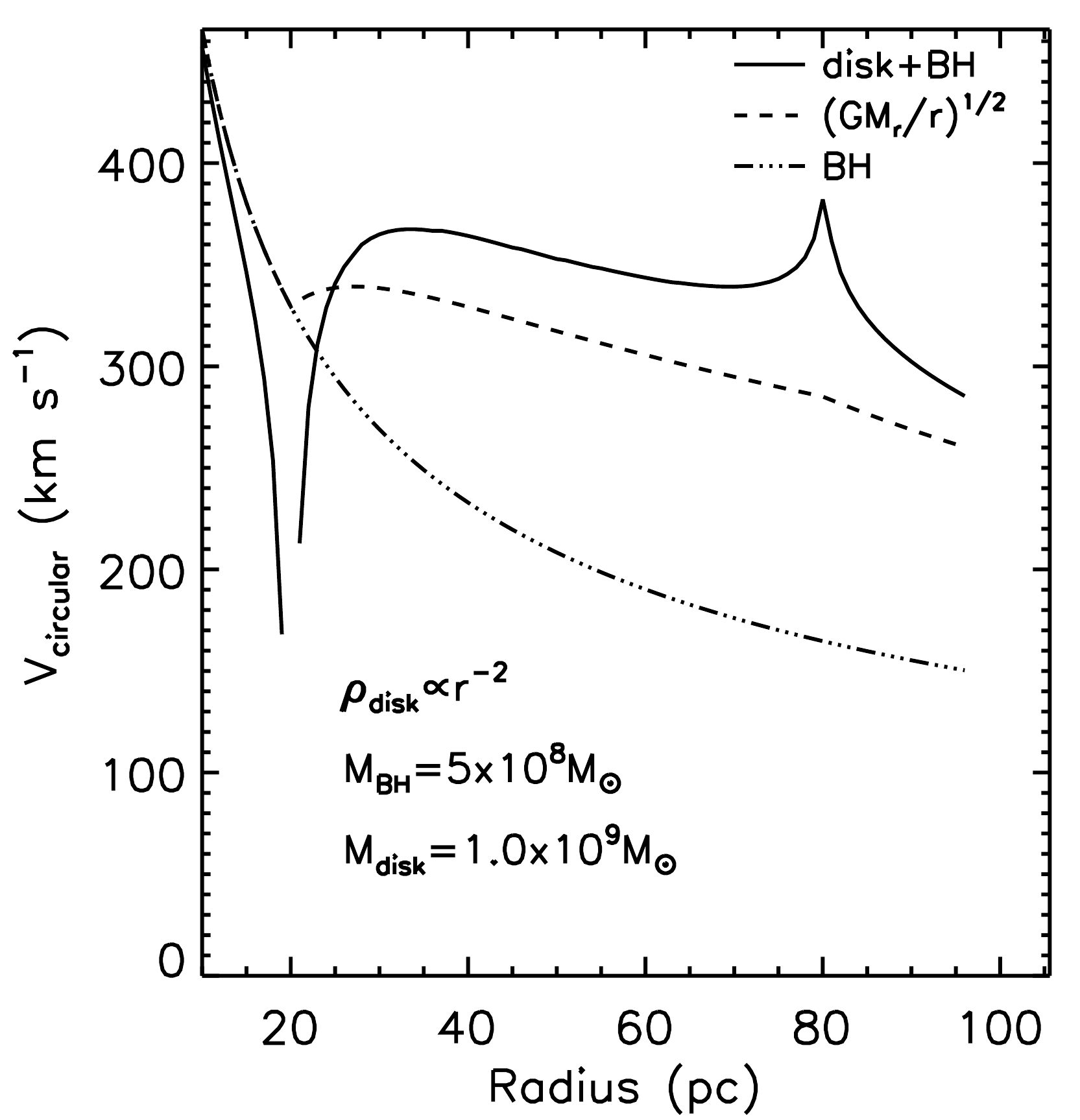}
\caption{The circular velocities are shown for a mass distribution including a central point-mass (BH) of $5\times10^8$ \msun , 
and disk of mass $10^9$ \msun ~distributed with an r$^{-2}$ density falloff between 20 and 80 pc. The solid curve shows the proper 
circular velocity obtained by integration over the disk mass distribution, and the dashed curve the spherical approximation, i.e. calculating $v_{circ} = (GM_r/r)^{1/2}$. The dot-dashed curve is the Keplerian rotation curve associated with the central point mass. The dip at 
20 pc is due to the net outward attraction in the inner regions of the disk; the cusp at 80 pc is due to there being no such outward 
attraction disk beyond 80 pc.}
\label{mod} 
\end{figure}

Lastly, we provide a note of caution on the comparison of dynamical masses with the masses determined from the 
CO line and dust continuum emission. If the dynamical mass is calculated in a simple spherical approximation from
$M_{dyn} = r V_{rot}^2 / G$, this mass estimate may be significantly different from that calculated from the observed kinematics assuming a disk mass distribution plus a central point mass. 

The mass distribution in the nuclei of Arp 220 is very likely disk-dominated and the spherical approximation is 
likely to incorrectly estimate the actual mass in such a configuration. We illustrate this with a simple model (similar to Arp 220 West): a central point mass of $5\times10^8$ \msun and a disk of $10^9$ \msun \, extending between 
r = 20 and 80 pc with surface density varying as $r^{-2}$ and constant thickness. Figure \ref{mod} shows the circular velocity with forces determined  
properly for a central mass plus the disk. The circular velocity estimated in the spherical approximation 
(from the enclosed mass at each radius) is also shown for comparison. 

The proper estimation (solid curve) and the spherical one (dashed curve)
have significantly different $v_{circ}$. The differences are due to the fact that in 
a thin disk configuration there is significant gravitational attraction in the outward direction by gas at larger radii. We defer a proper comparison of the emission-based mass estimates to a future paper, once we have full flux recovery of the CO and dust emission. 
The rotation curve shown in Figure \ref{mod} is somewhat higher than that in Table  \ref{models} -- but within the range of uncertainties 
due to the inclination angle. In Section \ref{rad_press}, we note an added uncertainty with the dynamical mass estimates -- that radiation pressure on the dust
may also provide substantial pressure support in the disks. 

\subsection{Physical Structure of the Nuclear Disks}\label{physics}

The high resolution imaging presented here provides strong confirmation for the existence of nuclear 
disk structures in both nuclei of Arp 220, as first suggested by \cite{sak99}. In the Eastern nucleus the gas 
and dust emissions are clearly elongated and this elongation aligns with the major axis of the CO velocity gradient. 
In Arp 220 West, the emission is not so clearly elongated but there is a strong gradient in the CO line velocity at PA $\simeq265$\deg.  
The lack of elongation in the emission distribution is likely due to the fact that the West nucleus is more compact and the 
kinematics suggest a more face-on inclination (inclination i = 30\deg, compared to 45\deg in the East nucleus -- Table \ref{models}). 

These nuclear disks are extraordinary structures with $\geq10^9$\msun within radii $<50 - 100$pc and the two nuclei are only
$\sim412$ pc apart. Here, we discuss the mean ISM properties in the nuclei, the physical structure of the disks, 
 maintenance of the disks and the energetics of the nuclear sources.  At this point, lacking the full datasets, our discussion is 
qualitative -- intended to be illustrative of the physical considerations. The present observations do not recover all of the line and continuum 
flux from the region since they include only long baseline data and only the $\lambda \sim 2.6$ mm observations; we anticipate 
a more thorough analysis when the low resolution observations are completed including CO 2-1 and 3-2 line emissions with complete flux recovery. 
The observations reported in this paper do 
not include short spacing data and hence they do not recover flux associated with the large scale structures $\gtrsim$0.5 \arcsec or 200 pc.

\cite{lon06} report VLBI at 18 cm wavelength and $\sim1$ pc resolution, detecting 49 point-like sources 
which they interpret as supernova remnants (SNR)\citep[see also][]{par07}. The sources are tightly clustered in the two nuclei -- with 75\% of the sources in two regions: $0.25\times0.15$\arcsec (West nucleus) and $0.3\times0.2$\arcsec (East nucleus). In both cases these rectangles are aligned with the PA derived above for the disks. The estimated 
supernova (SN) rate is $4\pm2$ yr$^{-1}$ based on the appearance of new SN between the epochs of the observations. The 
fact that 22 of the 49 SNR are in the East nucleus and 27 in the West nucleus strongly suggests that the total star formation rates are similar for the two nuclei, i.e.  
45\% vs 55\%. The fact that only 25\% of the 49 SNR are outside the two compact nuclear regions implies that relatively little (25\%) of the total SF occurs in the larger regions of Arp 220.

Lastly, we note one puzzle -- in the West nucleus, the kinematic modeling of the intensity and velocity distributions suggests a low inclination disk (i = 30\deg, Table \ref{models}), yet this is inconsistent with the elongated distribution of the SNRs (having a 2.5:1 major:minor axis ratio \citep{lon06}) for which one would infer i $\simeq 60$\deg for a thin disk. Two possible resolutions of this contradiction are: 1) either disk gas is more inclined than 30\deg but the North and South extents of the West nucleus gas are increased by minor axis 
outflows or there being a substantial thickness to the disk or 2) the SNR are preferentially along one axis within a low inclination disk. The latter might occur 
if the nuclear disk had a central bar in which the most recent SF was preferentially occurring. {\bf However, the rotation period of the disk is only a few Myr and it is 
unlikely that the massive stars formed in a bar would remain in a bar until the 
time that they undergo SN explosions.} Higher resolution imaging and kinematics is probably needed 
to address this inconsistency. 

\subsection{Summary of Observational Parameters}

For the purpose of the numerical evaluations in the discussion below we adopt approximate estimates for the masses, radii and luminosities 
for Arp 220 -- these are meant only for physical perspective in the discussion and are probably uncertain by factors of two or more.  
For the ISM masses, we adopt $\sim1.5\times10^9$ \msun \, for each of the Arp 220 disks 
with their radial extents being 80 and 130 pc for the West and East nuclear disks, respectively (Table \ref{models}) since the deconvolved radii are 74 and 11 pc (Table \ref{sizes}). The velocity dispersions 
derived from the line profile fitting are $\sigma_v = \Delta v_{FWHM}/2.3 \simeq 110$ and 50 \kms, respectively ($\sigma_{3d} = \sqrt 3 \sigma_{1d}$). 

In the West nucleus 
the central point-like concentration has a mass of between $8\times10^8$ and $3\times10^9$\msun ~as derived from the gas kinematics 
and the dust emission, respectively. Resolution of this discrepancy may be possible with the higher resolution 1.3 mm observations which will 
allow a better constraints on the kinematics and on the true dust temperatures in that source. Below, we adopt $10^9$ \msun ~for the unresolved mass 
within $R \leq 15$ for this extreme concentration. 

The total infrared luminosity of Arp 220 is $1.9\times10^{12}$ \lsun. Using the mid infrared, high resolution ($\sim0.4$\arcsec)
photometry from \cite[][Table 3]{soi99} we take the view that the majority of this luminosity arises from the two nuclei, apportioned 2/3 (West) and 1/3 (East) (i.e. 1.2 and 0.6$\times10^{12}$ \lsun). We caution that these estimates are approximate since the 3 - 24$\mu$m photometry  \citep{soi99} is not at the $\lambda \sim 70 \mu$m far infrared luminosity peak; also the distribution of SNR is more nearly equal for the two nuclei \citep{lon06,par07}.
Some of the overall luminosity is also likely to originate at larger radii; the above luminosities should therefore be taken as upper limits to the luminosity of each nucleus.  

\subsection{Disk Structure} \label{disk}

The formation of gaseous disk structures in the nuclei of merging galaxies is likely an inevitable result of the gas sinking dissipatively into the central regions
more rapidly than the main stellar component of the galaxies. This results in a gaseous bar leading the stellar bar (since the gas has sunk to smaller radii) and the 
stellar bar then exerts a backwards torque on the gaseous bar to further reduce its rotational angular momentum \citep[see][]{bar92,bar96}. The gas forms a rotating disk since vertical motions 
are efficiently damped out once the gas becomes concentrated, while the angular momentum is removed on a much longer timescale. 

The thickness of the disk is determined by equilibrium between the gravitational forces toward the midplane of the disk and the gas motions in the vertical direction. 
These gas motions will be damped if two parcels of gas collide; their bulk kinetic energy is then converted to thermal energy in shocks. The high density 
of the molecular gas ensures that the shocks radiate the post-shock thermal energy very efficiently. The net result is that the kinetic energy associated with vertical motions needed to maintain 
the disk thickness is radiated on a timescale similar to the collision time of the gas parcels. In order to maintain the thickness of the disk and its associated 
vertical motions on a longer timescale, constant replenishment of the turbulent energy is required. 

The typical timescale for collision of gas parcels and the dissipation of the turbulent motions is given by the vertical crossing time of the disk and the fraction of the disk area filled by gas parcels as viewed 
perpendicular to the disk: 
\begin{eqnarray}\label{dis}
 \tau_{\rm dis} &=& \tau_{\rm cross} / \rm f_a \nonumber \\
 &=& \rm H / (\sigma_v ~ f_a)  . 
\end{eqnarray}
 
\noindent Here H is the full thickness of the disk and $f_a$ is the areal covering factor of the disk. 

 \begin{figure*}[ht]
\epsscale{1}  
\plotone{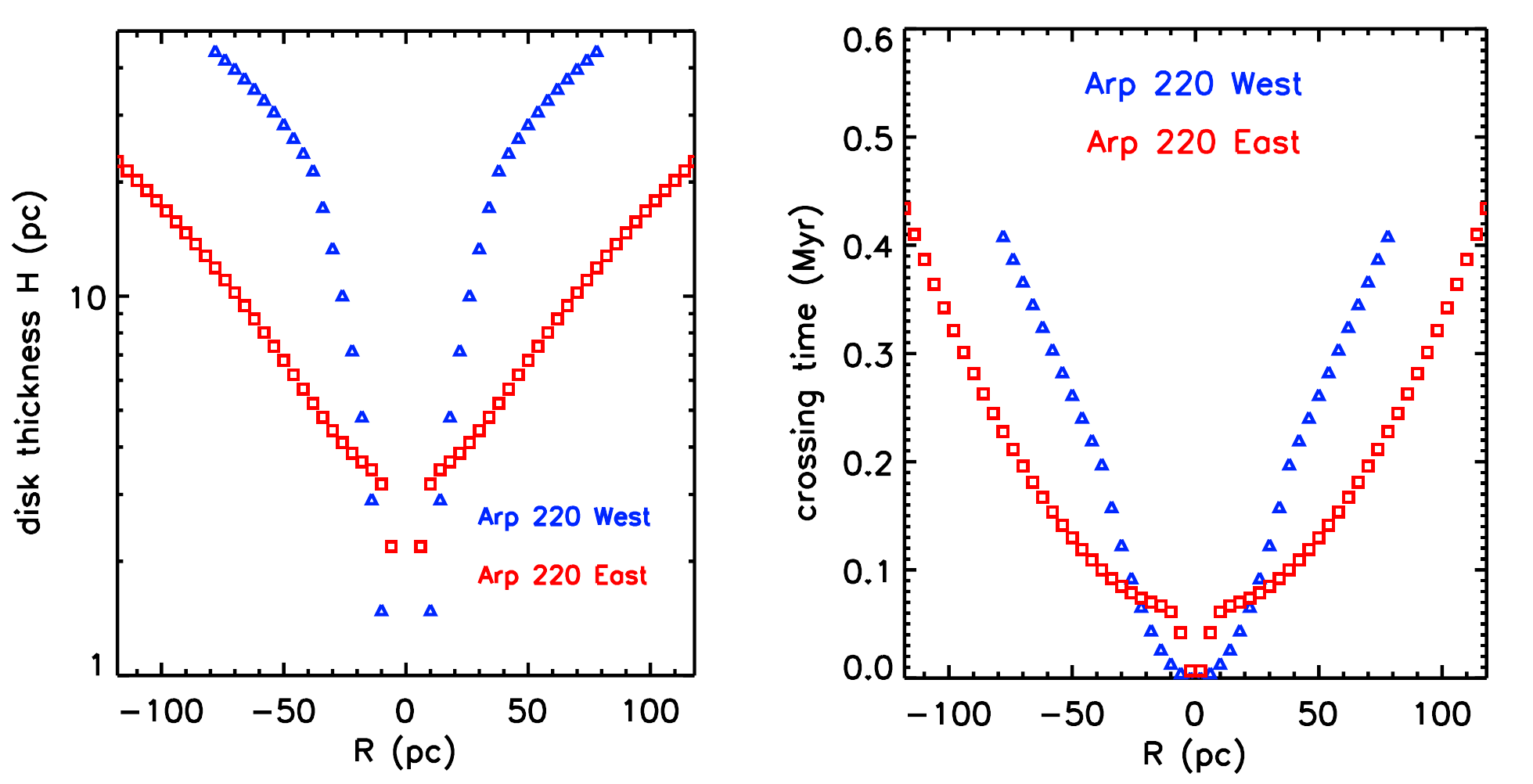}
\caption{{\bf Left panel:} The estimated scale heights of the West and East disks are shown based on Equation \ref{self} using the 
rotation curves shown in Figure \ref{rot_curve1} and velocity dispersions $\sigma_v(1d) = \Delta v (FWHM) / 2.3$ from Table \ref{models}, i.e. 1-d velocity dispersions were taken to be 109 and 52 \kms, respectively. {\bf Right panel:} The disk crossing timescale (H/$\sigma_v(1d)$) which is equivalent to the turbulent dissipation timescales calculated from Equation \ref{dis} with $f_a = 1$.}
\label{scale_height} 
\end{figure*}

\subsection{Disk Area Covering}

 The disk areal covering factor must be of order unity. The observed CO peak brightness temperature 
is comparable with the estimated dust brightness temperatures (200 K),and hence the dust temperature, since the dust is nearly optically thick. The CO peak brightness temperature in the West disk is 187 K (Table \ref{fits}) and the areal covering factor for the CO emitting gas 
must be $\sim1$. In the East disk, the CO peak is 175 K, implying a similarly high areal covering factor. % with $f_a \simeq 1$. Given the high gas densities, we also expect
%that the dust and gas temperatures will be coupled with $\rm T_K \sim T_d$. $T_d$ is determined by the radiative equilibrium of the dust. Since the 
%dust is very optically thick, we may use a blackbody approximation assuming a central source of luminosity L, to estimate $\rm T_d$ as a function of radius:
%\begin{eqnarray}
%\rm L &=& 4 \pi \rm \sigma T^4  ~~\rm yielding \nonumber \\
%\rm T &=& 194 \left( {\rm L \over{10^{12} \lsun }} \right) ^{1/4} \left( { \rm 20~pc \over{\rm R}} \right) ^{1/2}  .
%\end{eqnarray}\label{td}
%\noindent For a more distributed source such as a starburst spread over the disks, each local luminosity center will have a radial gradient in $T_d$ following Equation \ref{td}; in this 
%case there may be hotter grains at larger galactic radius. 

\subsection{Disk Thickness}

For the nuclear disks we may estimate the disk thickness assuming that the vertical distribution is determined by equilibrium between the gravitational 
force in the z-direction and the observed gas 1-d velocity dispersions ($\sigma_v(1d)$). We do this under the very simple assumption that the vertical 
velocity dispersion is constant with radius and that the disks are in equilibrium. Specifically, this assumes that outflow winds are not contributing to the vertical 
scale height.

We consider two cases: 1) where the gas disk mass surface density is much less than that of the stellar disk or spheroid, and 2) where the surface density is dominated by the gaseous 
disk (i.e. a fully self-gravitating disk). The former is appropriate for low-z galactic disks, where the ISM mass is typically only 5-10 \% of the 
stellar mass. The latter is likely to be most appropriate for high-z galaxies with a large gas-mass fractions or the gaseous nuclear disks 
as in Arp 220 where the gas has been preferentially funneled to the nucleus faster than the stars. 

In the non self-gravitating case, the vertical distribution of the gas will have density $\rho = \rho_0 \rm exp(-z^2/z_0^2)$.  If we define the disk thickness (H) as the full thickness at which the density has dropped by a factor 1/e, then H = 2 z$_0$. For a spherical 
distribution of stars, 
\begin{eqnarray}\label{td1}
\rm H(r) &=& 2\sqrt{2} \left( {\sigma_v(1d) \over{ V_{rot}}} \right) ~r ~.
\end{eqnarray}

\noindent For a typical low-z galaxy where the stellar distribution is disk-like, the vertical frequency $\nu_z / \Omega_{rot} \sim 3$ 
and the thickness is reduced by a corresponding factor $\sim3$. 

In a fully self-gravitating gas disk, the vertical distribution is $\rho = \rho_0 \rm sech^2(z/(2 z_0))$ with 
\begin{eqnarray}\label{td2}
\rm z_0(r) &=&    {\sigma_v(1d) \over{ \sqrt{8\pi G \rho_0}}}   \nonumber \\
&=&    {\sigma_v(1d)^2 \over{ 2\pi G \Sigma_{disk} }} 
\end{eqnarray}

\noindent where $\Sigma_{disk}$ is the mass surface density of the gas disk. For a Mestel disk, $\Sigma_{disk} = V_{rot}^2/(2\pi Gr)$ and therefore,
\begin{eqnarray}\label{td2}
\rm z_0(r) &=&    \left( {\sigma_v(1d) \over{ V_{rot}}}\right)^2 r ~~.
\end{eqnarray}

\noindent In this the case, the equivalent full thickness at which the density has dropped by 1/e is $H(r) \simeq 4.34 ~z_0$, i.e. 
\begin{eqnarray}\label{self}
\rm H(r) &=&   4.34~\left( {\sigma_v(1d) \over{ V_{rot}}}\right)^2 r ~~. 
\end{eqnarray}

%z$_0$ = the and the ratio of the scale height to the radius is given by $\rm H/R \simeq \sigma_v/ V_{rot}$ where $\rm V_{rot}$ is the rotational velocity as a function of radius. A similar scale height relation is obtained if the disk mass is dominated by a central point mass. On the other hand, if the disk mass dominates and the disk is therefore fully self-gravitating, 
%the vertical distribution is $\rho = \rho_0 \rm sech^2 (\rm z/H)$ and $\rm H/R \simeq \sigma_v^2/\rm V_{rot}^2$. [A continuous interpolation treatment of these cases is provided in \citep{ber99}]

Figure \ref{scale_height} shows the disk scale thickness (H) variation as a function of r for the West and East disks (Table \ref{models}),
indicating H ranging from 1 to 30 pc at r $<$ 50 pc. The disk crossing times (and thus the
turbulent dissipation timescales) shown in the right panel of Figure \ref{scale_height} are $10^5$ to $4\times10^5$ yrs. 

In order to maintain the disk vertical structure and its observed velocity dispersion, it is required that the turbulent 
energy in the gas motions be replenished on a similar timescale -- the sources of this input might include: 1) starburst and AGN power and the momentum 
associated with their radiation; 2) the pressure support provided by radiation liberated 
by the turbulence dissipation and 3) the gravitational potential energy released as the gas accretes inward and the potential energy 
associated with the decaying orbits of the two nuclei. These are discussed and evaluated in Section \ref{energy}.

\subsection{Disk Gas Properties}

For the masses and radii given above we estimate the mean gas densities in the two disks. From the fitting of the line profiles 
across the two disks, the best fit model has approximately constant CO line emissivities as a function of radius outside the central R = 10 pc, with 
the West disk having a factor 2-3 higher mean emissivity compared to the East disk; on the other hand, the East disk has a larger radial extent. 
We adopt mean disk thickness in the z-direction of 10 pc. The mean H$_2$ density (not correcting for He) is then given by:

\begin{eqnarray}\label{density}
 <\rm n_{H_2}> &=& \rm M_{disk} / (4\pi \rm R^2 \rm ~H ~m_{H_2})  \nonumber \\
 \rm <n_{H_2}> &=& 1.7\times10^4 \left( {\rm M_{disk} \over{10^{9} \msun }} \right) \times \nonumber \\
  && \left( { \rm 10~pc \over{\rm H_{FWHM}}} \right) \left( { \rm 100~pc \over{\rm R_{disk}}} \right) ^{2}  \rm cm^{-3}.
\end{eqnarray}

For the mean thickness of 10 pc and density $n_{H_2} \simeq 1.7\times10^4$ cm$^{-3}$, the column density through the disk is $N_{H_2} = 5.2\times10^{23}$ cm$^{-2}$. 
Assuming a standard Galactic ISM gas-to-dust ratio $N_{H_2} / A_V = 1.0\times10^{21}$ cm$^{-2}$ mag$^{-1}$, we find a typical extinction A$_V = 520$ mag. 
The disks will thus be optically thick to any NIR-UV radiation emitted by young stars formed within the disks. 

One final question is the structure of the molecular gas disk -- is it composed of a smooth gas distribution or discrete clouds (e.g. GMCs, or smaller clouds)? 
If the observed velocity dispersion $\sigma = 50 - 100$ \kms ~is due to internal motions within individual clouds and if these clouds are gravitationally bound, 
 virialized and have uniform density, then their masses are $M_{\rm cloud} = 5/3~R \sigma_v (\rm internal)^2 / G$. If the 3-d $\sigma_v (\rm internal) = 100$ \kms~ and one requires that 
 their density be twice the mean disk density ($<\rm n_{\rm H_2}> = 1.7\times10^4 \rm ~\rm cm^{-3}$), their masses must be $\sim7\times10^7$ \msun with radii = 18 pc (the resolution of the observations). However, the area covering factor 
 for the disk would then be inadequate ($f_a\sim0.1$) (and their size would be larger than the disk thickness).  
 
 For higher density self-gravitating clouds, the area covering factor is even lower. Lower density self-gravitating clouds would be larger, have a more appropriate area covering factor, but then their density contrast would be sufficiently small that one might as well think of the disks as continuous, rather than a cloudy medium. We are thus led to the conclusion that the gas cannot be in discrete clouds but must be a fairly continuous medium (albeit with some clumping) distributed over the disk area.
 
To summarize --~1) If the gas is indeed relatively smooth (i.e. not in discrete clouds as in the Galactic disk), it must still have highly supersonic 
 motions with Mach number $>100$. 2) If these motions are not in the form of smoothly varying flows, then the turbulent kinetic energy dissipates on a disk crossing time $\sim10^5$ yrs (Figure \ref{scale_height}-right). 3) The turbulent kinetic energy must then be replenished within the dissipation timescale 
 in order for the structure to last at least one rotation period of the disk. If the disk does not last this minimum time period, we should not expect to 
 see a disk structure with the high velocity dispersion observed. 4) Alternatively, if the motions are in an ordered flow (and therefore less-dissipative), 
 the scale over which the velocity field varies must be smaller than the spatial resolution of our observations $\sim30$ pc.  
 
 A hybrid picture -- clouds of dense gas (with low areal covering factor) embedded in a diffuse molecular medium of low column but high enough areal covering factor to account for the CO brightness temperatures -- might also be viable. Most of the 
 mass would have to be in the compact, high density clouds to avoid having the kinetic energy associated with their bulk motion rapidly dissipated
 by friction as they move through the diffuse medium. In this case, a large reservoir of kinetic energy is stored in the compact clouds which 
 stir up motions in the diffuse medium without dissipating significant kinetic energy. And of course, such clumps might not be long-lived, but rather being formed and then destroyed. Discriminating this hybrid picture from a more continuous 
 model will require high resolution measurements of the brightness temperatures for high excitation emission lines, to assess the areal covering factor 
 of the dense component and rare isotopes of CO to constrain the line optical depths.

\subsection{Energy and Momentum Considerations}\label{energy}

There are several sources of energy input to the gas which might replenish whatever losses there are in shock fronts: the radiative and mechanical energy and momentum 
input from the starburst, the gravitational potential energy resulting from shrinking of the two disks, and the final merging of the galactic nuclei. Taking a characteristic timescale
of 1 Myr (which is an upper limit to the dissipation timescale if the motions are dissipative), we may compare these sources. 

The total energy in the radiation 
field associated with L = $10^{12}$ \lsun~ integrated over 1 Myr is $1.1\times10^{59}$ ergs. The total energy contained in the motions associated with the observed velocity 
dispersion $\sigma_v(1d) = 100$ \kms is $3/2 (M \sigma_v(1d)^2) = 4.5\times10^{56}$ ergs for $M_{ISM} = 1.5\times10^9$ \msun (the 3 is to convert to 3-d velocity dispersion). The potential energy associated with each disk and with the binary merging galactic nuclei 
is larger than that calculated above for the dispersive gas motions, by a factor (V/$\sigma_v(\rm3d))^2 \gtrsim 1$, where V is the disk rotation velocity or the relative velocity of the nuclei. 

Although the energy in the radiation field is vastly larger than that in the gas motions, it is more appropriate to compare the radiation momentum with that in the gas, since the gas motions are more likely to be driven by the radiation pressure than by expansion of hot gas heated by the radiation field. 

The radiative momentum input (L/c per unit time) integrated over 1 Myr is $3.7\times10^{48}$ gr cm sec$^{-1}$; the momentum in the gas is $3\times10^{49}$ gr cm sec$^{-1}$, i.e. an order of magnitude larger. The momentum input over the same time period from supernovae at a rate $\sim$4 yr$^{-1}$ \citep{lon06} (assuming 5 \msun ~of ejecta in the momentum conserving phase starting at $v_{exp} \lesssim 1000$ \kms), is $\lesssim 4\times10^{48}$ gr cm sec$^{-1}$. 
In summary, if the gas motions are dissipated on a disk crossing timescale, there is no obvious source for their replenishment other than the large scale
gravitational field of the galaxy. 

\subsection{A Semi-Coherent Disk Wave Pattern}

In the last section, we found that there are no known localized sources of energy to replace that which is dissipated on a disk crossing timescale. Here we explore the 
possibility that the observed velocity dispersion is due to "coherent" flows or waves within the disks. The 
flows must be semi-coherent to reduce dissipation. The scale over 
which these flows must span the full range of velocities ($\simeq2\sigma_v$) must be significantly smaller than the observational resolution of 37 pc. Clearly, this will not avoid all dissipation, since there will be some regions where the flow still has shocks, 
but in the absence of dispersive motions throughout, the overall dissipation might be reduced an order of magnitude. One possible scenario could be wave motions in the disks generated by the tidal field of the two merging 
galaxy nuclei -- a dumbbell-like potential. The characteristic scale of such gravitational perturbations is $\gtrsim100$ pc, i.e. much larger than the required 30 pc scale. On the other hand, one expects that the energy input at long wavelengths should decay to smaller wavelengths, as in a turbulent cascade. 

A gas parcel of unit mass moving in the z direction perpendicular to the disk will execute harmonic motion relative to the mid plane with the restoring 
force due to the central point mass and the disk mass at lower z height:
\begin{eqnarray}\label{}
\rm F_{central} &=& \rm -G {M_r \over {\left( r^2+z^2 \right)^{3/2}}}z \simeq -G {M_r \over { r^{3}}}z  ~~~ \rm and \nonumber \\
\rm F_{disk} &=& \rm -4\pi G \rho z  \nonumber 
\end{eqnarray}

\noindent where the disk is assumed to have constant mass density $\rho$ over scale height H at radius r, and $M_r$ is the mass interior to r (disk and central mass). The scale length H for the harmonic motion is then 
\begin{eqnarray}\label{}
\rm H^2 &=& { 2 \sigma_v^2 \over{\omega_c^2 + \omega_d^2 }} ~~~ \rm where \nonumber \\
\omega_c^2 &=& {\rm GM_r \over{r^3}} \rm ~~~~ and ~~~~~\omega_d^2 = 4 \pi G \rho .  
\end{eqnarray}

%Writing the angular frequencies in terms of $V_{rot}(r)$, the vertical scale length is 
%\begin{eqnarray}
%\rm H/R &\simeq& .... \\
%\end{eqnarray}\label{}

\noindent These waves would be analogous to gravity waves -- possibly generated by the tightly 
wrapped spiral pattern excited in late stage mergers by the companion nucleus. 

\section{Radiation Pressure Support}\label{rad_press}

As noted in Section \ref{dyn_mass}, it might appear difficult to reconcile the mass estimates for ISM within the disks (as well as any 
central components such as a black hole or compact dust source) with the observed rotation curves assuming circular motions. However, 
the high opacity of the dust will mean that there is substantial absorption and reemission of the infrared radiation and, therefore, 
pressure support from the radiation field, since the dust is optically thick, even at far infrared wavelengths \citep{sco03,tho05,tho15}. 

Assuming the Mestel disk, with L$_r$ and M$_r$ being the luminosity and mass as a function of radius in the disk, then the 
condition that inward gravitational forces are balanced by the outward radiation pressure is
\begin{eqnarray}\label{rad1}
{ \rm L_r \kappa \rho \over{c 2 \pi r H}} &\simeq& { \rm G M_r \rho \over{r^2}} ~~~ \rm ,~implying \nonumber \\
{ \rm L_r \over{M_r}} &=& {\rm c 2 \pi H G \over{r \kappa}}
\end{eqnarray}

\noindent where $\kappa$ is the radiation absorption coefficient per unit mass weighted to the effective wavelength 
of the infrared radiation field at each radius. Using Equation \ref{self} for the ratio H / r, Equation \ref{rad1} becomes
\begin{eqnarray}\label{rad5}
{ \rm L_r \over{M_r}} &=& {\rm 8.6 \pi c \over{\kappa}} G \left({\sigma_v \over{V_{rot}}}\right)^2 ~.
\end{eqnarray}

Lastly, we require $\kappa$ at the wavelengths of the far infrared radiation in the disks. For the standard ISM gas-to-dust 
ratio $N_H / A_V = 2\times10^{21}$ cm$^{-2}$, $\kappa_{V} \simeq 312$ cm$^{2}$ gr$^{-1}$ and for $\kappa_{IR}/\kappa_{V} \sim 1/30$ \citep{pol94,sco13d,tho15}, 
 $\kappa_{IR} \simeq 10$ cm$^2$ gr$^{-1}$. Equation \ref{rad5} then becomes 
 \begin{eqnarray}\label{rad6} 
{ \rm L_r \over{M_r}} &=& 312 \left({\sigma_v / V_{rot} \over{1 / 3}}\right)^2  {\lsun \over{\msun}} .
\end{eqnarray} 

For a disk of luminosity $\sim5\times10^{11}$ \lsun ~and mass of $\sim10^9$ \msun ~as in Arp 220, the luminosity-to-mass ratio 
is 500 \lsun / \msun. It is therefore clear that radiation on the dust in the central regions of Arp 220 
is likely to be very important in providing significant pressure support against gravity. Of course this also 
implies that estimates of the dynamical mass from the observed kinematics may substantially underestimate the 
enclosed masses. In this case, the radiation is providing significant pressure support and thus the required circular velocities 
are smaller.  

The analysis above did not allow for the higher escape probability of the IR radiation 
in the z-direction (as opposed to the radial direction). However, the disks are also quite optically thick 
in the z-direction at the wavelengths of the $\sim $100 -- 200 K radiation field, so the escape probability perpendicular to the disks is low.

\section{Summary Discussion}

These CO (1-0) and dust continuum observations with ALMA at 90 mas resolution clearly resolve the two nuclei in Arp 220. 
The two nuclei have rotating disks of radii 74 (West) and 111 pc (East) (Table \ref{sizes}). The West nucleus also has a massive, unresolved 
dust emission source in its center which is optically thick at $\lambda = 2.6$ mm; the column density of gas in this compact 
source is $\sim2\times10^{26}$ cm$^{-2}$, assuming the dust has normal ISM dust opacity and abundance. This column 
corresponds to an incredible A$_V = 2\times10^5$ mag and 900 gr cm$^{-2}$ -- equivalent to 1 ft thick wall of gold!

Modeling the observed CO emission with rotating 2D disks in each nucleus yields acceptable fits observed line profiles. The 
kinematics of the West nucleus suggest a central point mass of $\sim8\times10^8$ \msun (an extremely compact gas concentration of a central black hole). Typical velocity 
dispersions in the gas are $\sigma_v / V_{rot} \sim 3$, indicating disk thicknesses between 1 and 30 pc over the range of radii. 
The calculated timescales for dissipation of the motions represented in the velocity dispersion is $\sim10^5$ yrs if these motions are turbulent. 
Since there does not appear to be a source for renewal of the turbulent energy at this short timescale, we suggest that there 
may be coherent wave-like motions in the disks. Such motions might be associated with the rapid wrapping of a spiral pattern 
generated by the close proximity of the nuclei. 

One of the most impressive aspects of the structure in the nuclei of Arp 220 is its symmetry and regularity. Despite the fact
that this is a very late stage galaxy merger with nuclei only 400 pc apart, the two disks are symmetric to a factor 
two in brightness and their kinematics can be modeled reasonably well with axisymmetric rotation curves. This symmetry 
likely implies that each disk has rotated several revolutions within the time over which they significantly 
change their separation. That is they have had time to relax during the merging process. 
For both disks, the rotation velocity is 300 \kms ~at 50 pc radius, so this rotation period 
is 1 Myr; the observed symmetry therefore implies a merging timescale $\gtrsim$3 - 5 Myr. 

The above is not to say there are no asymmetries. On the West 
nucleus there is a deficiency in the emission to the SW of the nucleus, i.e. a break in the ring of 
peak emission encircling the core. In addition there is also elongation of blue shifted emission to the south of the nucleus
(see Figure \ref{line_fig}). And in the East nucleus, the line emission peak is displaced $\sim0.07$\arcsec NE 
of the continuum peak ($\Delta \alpha, \Delta \delta = 0,0$ in Figure \ref{line_fig}). 

To put the Arp 220 structures in perspective, it is useful to compare with the dense gas structures in the 
center of the Milky Way. There the most massive gas concentrations are the Sgr B2 and Sgr A molecular clouds (each several $10^6$ \msun~ and size $\sim30$ pc) are separated by over 150 pc. The total molecular gas mass inside 300 pc Galactic radius is 
$\sim10^8$ \msun; thus only a small fraction of the area in the nuclear disk of the Milky Way is covered by dense gas, in contrast to Arp 220 where the areal filling factors are $\sim$1.
 
 \medskip
There are several topics we have not discussed here since they are more appropriately treated once we have the 
full CO (2-1) and (3-2) datasets with both high and low resolution imaging for full flux recovery. These include:
a comprehensive assessment and comparison of the masses (derived from line and dust emission and from the kinematics); 
the environs and interaction of the two nuclei; and an excitation analysis to estimate abundances, densities and temperatures 
in the gas. All are deferred to our later paper.

%They are not necessary for the analysis of the nuclei which is presented here. However, they are
%needed for a complete picture of the environs and interaction of the two nuclei and for comprehensive mass assessments.  Another objective of this 
%project is to thoroughly evaluate the use of CO lines and dust continuum for estimation of gas and ISM masses. This topic is deferred 
%until the low resolution imaging with full flux recovery are available along with the higher CO transitions at high resolution. 

\acknowledgments
%\begin{quote}
We thank the referee for a very thorough review of the manuscript and useful suggestions for clarification. We thank Zara Scoville for proof reading the manuscript. 

This paper makes use of the following ALMA data:  $\dataset[\rm ADS/JAO.ALMA#2015.1.00113.S]{\rm https://almascience.nrao.edu/aq/?project\_code=2015.1.00113.S}$.  ALMA is a partnership of ESO (representing its member states), NSF (USA) and NINS (Japan), together with NRC (Canada) and NSC and ASIAA (Taiwan), in cooperation with the Republic of Chile. The Joint ALMA Observatory is operated by ESO, AUI/NRAO and NAOJ. 
This work was done in part at the Aspen Center for Physics, which is supported by National Science Foundation grant PHY-1066293. TAT is supported by NSF Grant \#1516967. TAT thanks the Simons Foundation and organizers Juna Kollmeier and Andrew Benson for support for the {\it Galactic Winds: Beyond Phenomenology} symposium series.%\end{quote}

%\vfill
%\clearpage

\bibliography{arp220}{}

\appendix

\section{Observed and Model Fit CO Spectra}\label{co_spectra}

\begin{figure*}[ht]
\epsscale{1}  
\plotone{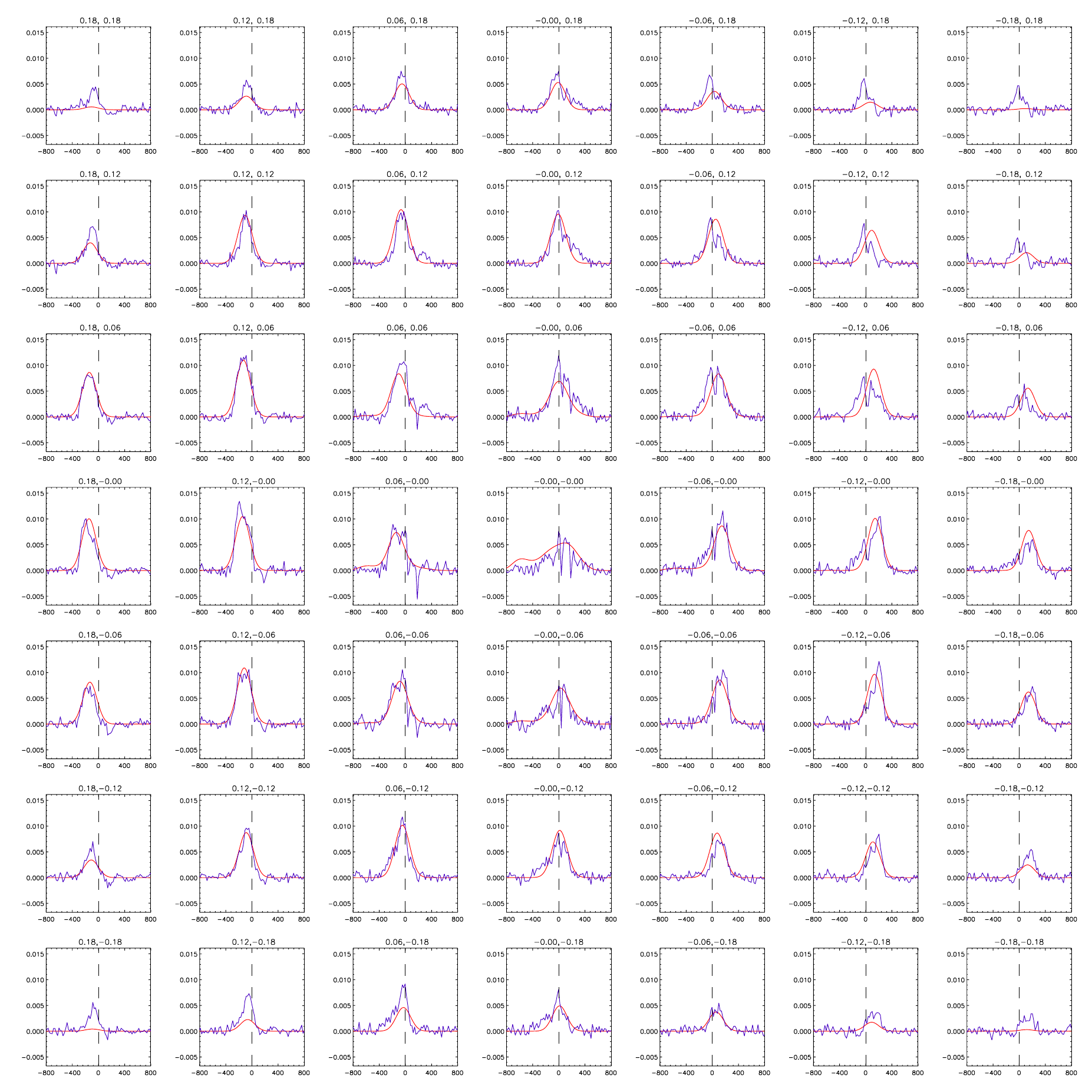}
\caption{The observed and model spectra (blue and red curves, respectively) are shown for a 60 mas grid centered on Arp 220 West. The legend above each spectrum provides the angular offset in arc sec and the velocity scale is relative to the West nucleus systemic velocity $\rm v_{radio} = 5337$ \kms.}
\label{west_spectra} 
\end{figure*}

\begin{figure*}[ht]
\epsscale{1}  
\plotone{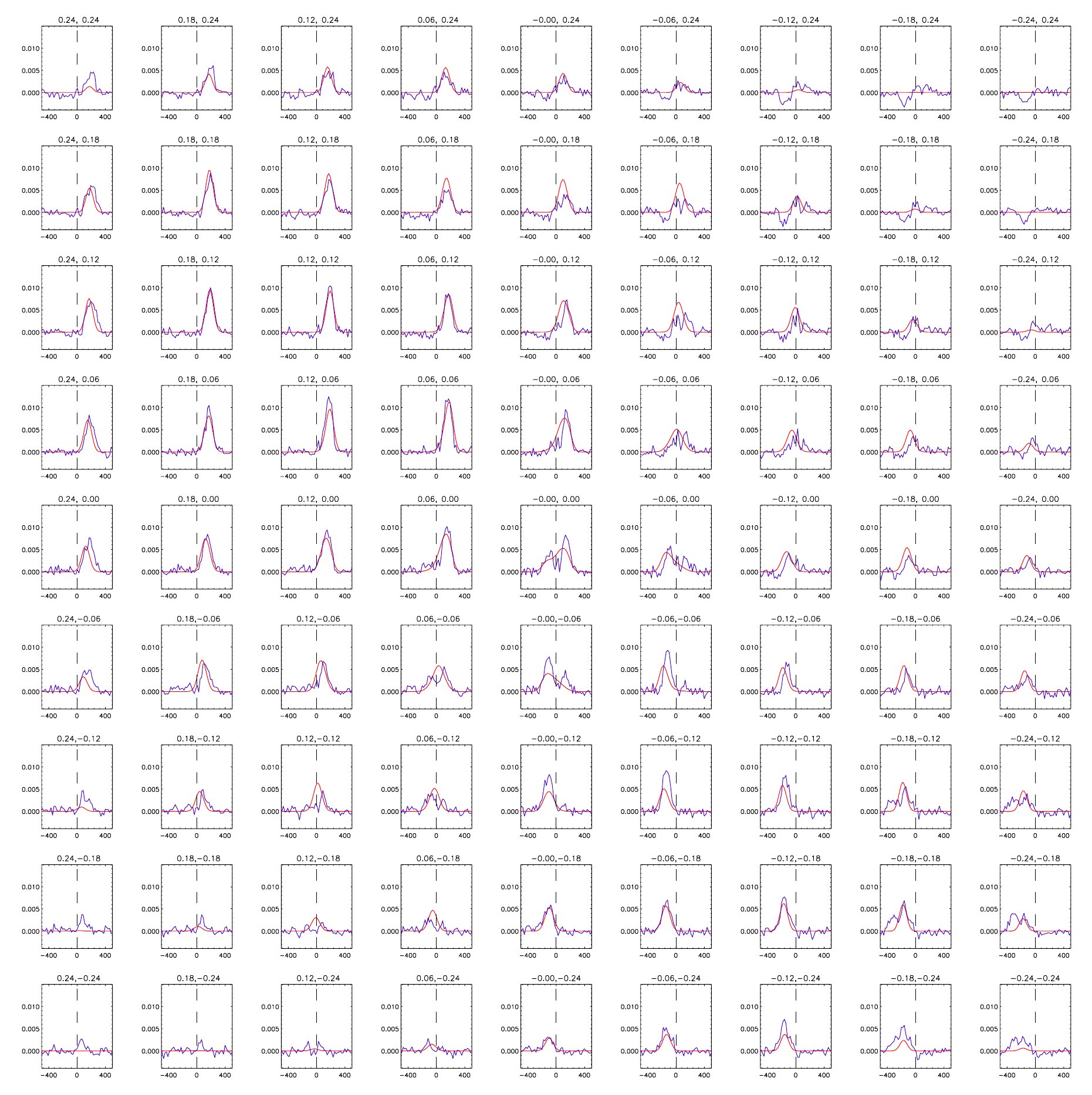}
\caption{The observed and model spectra (blue and red curves, respectively) are shown for a 60 mas grid centered on Arp 220 East. The legend above each spectrum provides the angular offset in arcsec and the velocity scale is relative to the East nucleus systemic velocity $\rm v_{radio} = 5431$ \kms.}
\label{east_spectra} 
\end{figure*}

In Figures \ref{west_spectra} and \ref{east_spectra} we show the observed and model CO spectra sampled on a 60 mas grid for the 
West and East nuclei. The spectra are shown here so that the reader can see that even at 90 mas resolution the CO emission 
has a large velocity dispersion $\sigma_v = 110$ and 50 \kms for the West and East nuclei. The appropriateness of the model fit 
parameters (see Section \ref{modeling}) can be judged by comparison of the blue (observed) and model (red) spectra. The model was 
a very simple semi-axisymmetric emissivity distribution with a parametrized rotation curve and a single velocity dispersion. 

\end{document}